\documentclass[Journal, comsoc]{IEEEtran}
\IEEEoverridecommandlockouts
\usepackage{cite}
\usepackage{amsmath,amssymb,amsfonts}
\usepackage{textcomp}
\usepackage{xcolor}
\def\BibTeX{{\rm B\kern-.05em{\sc i\kern-.025em b}\kern-.08em
    T\kern-.1667em\lower.7ex\hbox{E}\kern-.125emX}}

\usepackage{amssymb}
\usepackage{stackrel}
\usepackage{graphicx}
\usepackage{cite}
\usepackage[top=0.75in,bottom=1in,left=0.625in,right=0.625in]{geometry}
\usepackage{tikz}
\usepackage{hyperref}
\usetikzlibrary{%
	calc,
	arrows,%
	shapes.misc,
	shapes.arrows,%
	chains,%
	matrix,%
	positioning,
	scopes,%
	decorations.pathmorphing,
	shadows,%
	decorations.text
}
\usepackage{longtable}
\newtheorem{lemma}{Lemma}
\newtheorem{theorem}{Theorem}

\begin{document}
	\title{Theoretical Concept Study of Cooperative Abnormality Detection and Localization in Fluidic-Medium Molecular Communication}
	
	\author
	{Ladan~Khaloopour, Mahtab~Mirmohseni, \IEEEmembership{Senior Member, IEEE}, \\and Masoumeh~Nasiri-Kenari, \IEEEmembership{Senior Member, IEEE}	
		\thanks{This work was supported by the Iran National Science Foundation (INSF) Research Grant on Nano-Network Communications.
			This paper was presented in part in the 2020 Iran Workshop on Communication and Information Theory (IWCIT) \cite{khaloopour2020iwcit}, \url{https://ieeexplore.ieee.org/abstract/document/9163534}.}
		\thanks{The authors are with the Department of Electrical
			Engineering, Sharif University of Technology, Tehran,
			Iran (e-mail: ladan.khaloopour@ee.sharif.edu; mirmohseni@sharif.edu;
			mnasiri@sharif.edu).}
	}
	
		\maketitle
		
				
				\begin{abstract}
					In this paper, we propose a theoretical framework for \emph{cooperative} abnormality \emph{detection} and \emph{localization} systems by exploiting molecular communication setup. The system consists of mobile sensors in a fluidic medium,
					which are injected into the medium to search the environment for abnormality. Some fusion centers (FC) are placed at specific locations in the medium, which absorb all sensors arrived at their locations, and by observing its state, each FC decides on the abnormality existence and/or its location.
					To reduce the effects of sensor imperfection, we propose a scheme where the sensors release some molecules (\emph{i.e.,} markers) into the medium after they sense an abnormality.
					If the goal is abnormality detection, the released molecules are used to cooperatively activate other sensors.
					If the goal is abnormality localization, the released molecules are used by the FCs to determine the location. In our model, both sensors' imperfection and markers background noise are taken into account. For the detection phase, we consider two sensor types based on their activation strategy by markers. 
					To make the analysis tractable, we assume some ideal assumptions for the sensors' model.
					We investigate the related binary hypothesis testing problem and obtain the probabilities of false alarm and miss-detection. It is shown that using sensors with the ability of cooperatively activating each other can significantly improve the detection performance in terms of probability of error.
					For the localization phase, we consider two types of FCs based on their capability in reading sensors' storage levels. We study their performance and obtain the optimal and sub-optimal decision schemes and also the probability of localization error for both perfect and imperfect sensing regimes.
				\end{abstract}
				

	\section{Introduction}
	\IEEEPARstart{M}{olecular} communication (MC) is a new communication paradigm, in which the molecules or ions are used as information carriers.
	MC has different applications in both micro-scale and macro-scale environments, where the distances between the transmitter and the receiver are up to a few micro-meter and meters, respectively.
	MC can be employed for healthcare applications in bio-nano-environments \cite{felicetti2014molecular, mosayebi2018early, nakano2016leader, okaie2018leader}, macro-scale applications such as environmental monitoring \cite{akyildiz2011nanonetworks, roberts2002turbulent}, industrial applications and pollution detection \cite{NakanoBook, farsad2013tabletop}.
	
	In fluidic medium environments, such as pipelines, different abnormalities may occur. It is important and in some cases very critical to detect and localize these abnormalities. The main challenge is the difficulty of physical access to these environments. As a solution, we propose a theoretical framework for an MC-based system using mobile sensors to achieve both goals of abnormality detection and localization.
	Our goal is to theoretically analyze the possibility of using MC setup in detection/localization applications and compare different feasible setups according to the derived performance metrics. While the detection using MC may have many usecases in practice, future experimental validations are necessary that may benefit from the proposed framework.
	The prospective practical usecases of the proposed framework are targeted drug delivery and target (or abnormality) detection (leakage, pressure drop, pollutant entry) in pipelines. 
	In the latter case, traditional solutions are using acoustic and radar waves, which face some difficulties such as wave reflections, pipes materials, pipe depth in the ground, and liquid fluids \cite{Zhu2017, hunaidi1998ground, lockwood2003study}.

	\subsection{Related Works}
	
	The detection problem in an MC setup has been studied in existing works, where the target is usually an abnormality, and mostly, the sensors are used for detection. The sensors can be fixed or mobile (moving in the environment and searching for abnormality). Two types of targets are considered. In the first type, the target releases some molecules (known as markers) into the environment, which are used by the sensors for detection \cite{mosayebi2018early} and by a receiver to localize the location of molecule (marker) source \cite{qiu2015long, gulec2020fluid}.
	In \cite{mosayebi2018early}, the mobile bio-sensors are injected into a fluidic medium (\emph{i.e.}, blood vessel) and move to the proximity of the target (\emph{i.e.}, cancer cells) to detect higher concentration of markers released by the cancer cells. The sensors do not release molecules themselves (i.e., no cooperative activation). It is shown that the system performance at the presence of mobile sensors is improved compared with the static sensors. In \cite{mosayebi2018early}, the abnormality localization is not considered. In \cite{qiu2015long}, a receiver robot is used to localize the transmitter of molecules in an ocean.
	The robot uses Rosenbrock Gradient search algorithm to reach the transmitter by sensing its released molecules.
	The authors in \cite{gulec2020fluid} consider an experimental setup where the transmitter releases droplets into the air, which are received by a sensor. The mechanics of droplets in the air make it difficult to derive an analytical model. Thus, data analysis and machine learning methods are proposed for distance estimation.
	The authors in \cite{turan2018} focus on abnormality localization in a vessel-like environment with Poiseuille flow, using ring-shaped receivers. The localization is based on the mean peak time of the observed molecules. The peak time depends on the molecules releasing shape by the abnormality in time, which is not discussed. On the other hand, obtaining the mean peak time requires to repeat the measurements, but it is seems difficult to control the abnormality to repeat the releasing process.
	
	In the second type of targets, no molecule is released from the target and sensors release molecules after detecting abnormality \cite{okaie2018leader, mosayebi2017cooperative, nafise2020abnormality,  nakano2016leader, nakano2016performance, okaie2014cooperative}.
	Here, the sensors can follow one of these two scenarios: (\emph{i}) they only detect and/or localize the abnormality \cite{mosayebi2017cooperative, nafise2020abnormality, felicetti2014molecular}; (\emph{ii}) they move to the target point to perform an action \cite{nakano2016leader, nakano2016performance, okaie2018leader, okaie2014cooperative}.
	In works on scenario (\emph{i}), the goal is abnormality detection. In \cite{felicetti2014molecular}, a biological system is used for tumor detection in a blood vessel. Two types of mobile sensors are injected. The first type goes near the tumor cells and releases molecules, which propagate in the blood and may be received by the second type of sensors that relays the information to a desired recipient. \cite{mosayebi2017cooperative} uses several fixed and ideal (noise free) sensors for monitoring a tissue. Each sensor senses a part of tissue and cooperates with other sensors by sending molecules to a fusion center (FC) through a three dimensional (3-D) diffusion channel, if senses an abnormality. The focus of  \cite{nafise2020abnormality} is monitoring the abnormality propagation by determining the location and the time of the changes, using non-cooperative fixed sensors releasing different molecules to the FC, where the channel model is the same as \cite{mosayebi2017cooperative}.
	The scenario (\emph{ii}) is applicable for targeted drug delivery \cite{nakano2016leader, nakano2016performance, okaie2018leader, okaie2014cooperative}.
	In \cite{nakano2016leader, nakano2016performance}, a non-fluidic 2-D bounded area is considered and a system including two types of nano-machines is presented to detect a target and deliver the drug there. The first type nano-machines (leader bio-sensors) move in the environment to detect a target, then they release some markers there. The second type nano-machines (follower bio-sensors) carry the drug and move to the higher marker concentration points. Then, they release the drug at the target point. In \cite{nakano2016leader}, the movement of nano-machines is considered as a Weiner process, but the authors in \cite{nakano2016performance} consider a random-walk model. This system is extended in \cite{okaie2018leader}, where a new type of nano-machines is used to amplify the concentration of markers released by the first type of nano-machines, showing  performance improvement.
	The target tracking system is proposed in \cite{okaie2014cooperative}, where the target can move in a non-fluidic 2-D bounded environment and the bio-sensors have the ability of releasing and absorbing two types of repellents and attractants markers, in order to spread and guide other bio-sensors, respectively. The bio-sensors are bacterial chemotaxis, which have rotational diffusion mobility to track the target. The bio-sensors have a fixed speed, but the directions of their movements are changed based on the concentration of mentioned markers around them.
	\begin{figure*} [t]
		\centering
		\includegraphics[trim={0cm 4cm 0cm 6cm},clip, scale=0.5]{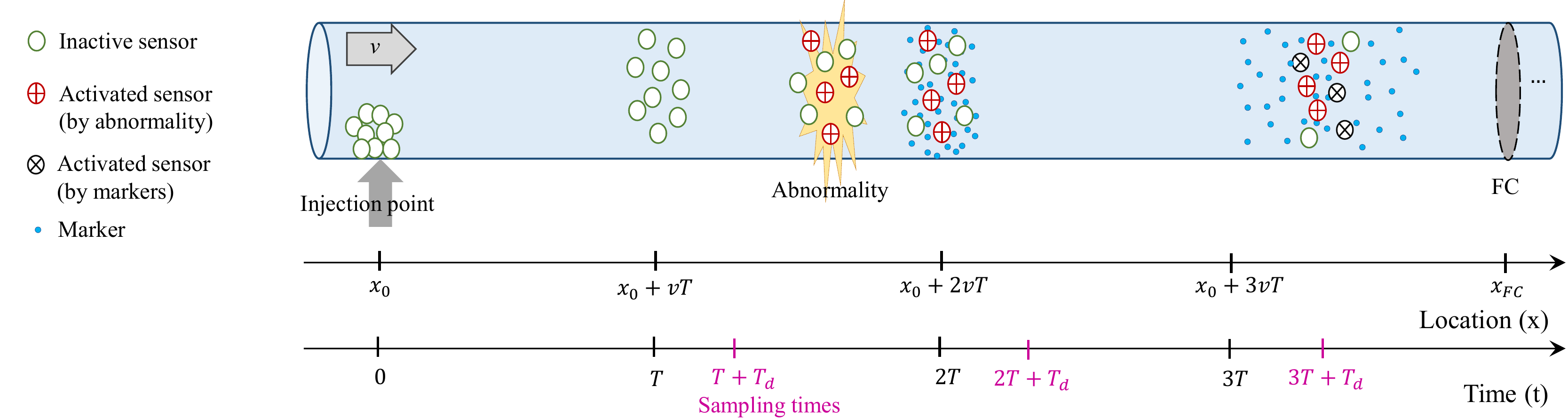}
		\caption{An arbitrary sensory region: inactive sensors are released from injection point $x_0$ and move with velocity of $v$. At the abnormality point, some of them are activated and then release markers. Some of inactive sensors are activated by sampling markers at the sampling times. Finally, the FC absorbs all sensors.}
		\label{FCs_regions}
		\vspace{0em}
	\end{figure*}
	
	
	\subsection{Our Contribution}
	In fluidic mediums such as pipelines, different abnormalities may occur, such as leakage, pressure drop, and pollutant entry. 
	In this paper, we propose and analyze a theoretical MC-based framework to employ the \emph{mobile cooperative} sensors for detection and/or localizing the abnormality in fluidic mediums. 
	To the best of our knowledge, the abnormality detection problem using cooperative mobile sensors in a diffusive fluidic MC setup has not been previously studied.
	We consider a cylindrical fluidic medium in a laminar flow condition, without any physical access to each point of it. We use mobile sensors to move in the medium and possibly reach the vicinity of abnormality. The sensors have the ability of sensing the abnormality and accordingly being activated \cite{nakano2016performance}. The medium is divided into some sensory regions. At the end of each region, an FC is located to absorb all of the sensors as they arrive, for example by a magnet \cite{Xie2012,Cohen2013}, and decide about the abnormality in its related region, using sensors activation states, similar to the FC used in \cite{mosayebi2018early}.
	Here, we have two challenges: the first one is sensors' imperfection, meaning that the sensors have miss-detection error, first studied in our earlier results \cite{khaloopour2020iwcit}. Our solution for this challenge is based on direct cooperation, where the sensors activated by sensing the abnormality release some molecules into the environment to activate the others. A similar sensors activation scheme by molecules is proposed in \cite{nakano2016performance}. These molecules are usually called \emph{markers}.
	The second challenge is localization, first studied in this paper, where the FC after detecting an abnormality in its region wishes to find its location. Our solution is to use a molecule releasing scheme to localize the abnormality.
	To detect the location, we attempt to find the time when an abnormality is detected by sensors, considering the fact that the location of the abnormality is proportional to the time passed since its detection\footnote{Because the sensors which detect the abnormality move with flow velocity.}.
	To respond to both of mentioned challenges, we exploit sensors equipped with storages to store molecules, which can be charged by a limited molecule production rate. This type of sensors is proposed in \cite{arjmandi2016ion} for natural cells and also is used in \cite{Bafghi2018, khaloopour2019adaptive}. These sensors, being completely charged at the time of injection into the medium, instantaneously release all stored molecules at the point where they sense an abnormality. Then, their storages are linearly charged until they reach the FC. Because of the linear production rate, the storage level of a sensor is proportional to the time duration between its activation and its absorption by the FC. Hereby, the FC can find the time that the molecules (markers) are released if it can read the levels of sensors' storages.
	After reading the states and storages levels of the sensors, the FC resets the sensors states, and after being fully charged, releases them to the next region.
	
	We exploit the theoretical models for considering the effects of sensors' imperfection and also marker background noise. The first one is when a sensor is activated because of wrong abnormality caused by the imperfect sensing operation, such as very small pollution concentration in the medium, or little pressure drop. We model this effect by a uniform sensor activation probability along the region. The second one is due to the existence of markers from other sources in the environment, which is modeled as a Poisson noise with a fixed rate as \cite{mosayebi2017cooperative}. The proof of marker noise distribution is explained in \cite{jamali2019channel}. In addition, we consider the probability that the sensor may be not activated encountering actual abnormality.
	
	For the goal of abnormality detection, we consider two types of sensors based on their activation strategies using the sampled markers. The first type is the memoryless sensor, which is activated based on the number of received markers in one sampling time, for example the PH-meter in \cite{khaloopour2019exp}, alcohol detectors in \cite{farsad2013tabletop}, and magnetic detector for nano-particles in \cite{Unterweger2018}, and the second type is the aggregate sensor that is activated based on the summation of received markers in all sampling times, before reaching the FC, for example the sensors used in \cite{mosayebi2018early}. Based on the considered theoretical sensor models, we characterize the system performance metrics analytically. First, knowing the number of sensors activated by sensing the abnormality, we derive the probabilities of false alarm and miss-detection and also the analytical probabilities of error, for two mentioned sensor types.
	We show that cooperation of sensors in activating each other significantly improves the system performance in terms of probability of error.

	For the goal of abnormality localization, we do not consider the cooperation among sensors (no sensor is activated by markers). Although simpler sensors are required here, we need more sophisticated FCs to decide on the location of detected abnormality through a multi-hypothesis testing problem.
	We consider two types\footnote{Another type of FC can also be considered, which has a simple structure and has not any knowledge about the sensors' storage. It can be shown that its performance is not acceptable, resulting in high probability of error.} of FCs for abnormality localization. Type-A FC can determine whether each sensors' storage is full or not when they reach the FC. But it cannot specify the sensors' storages levels. Thus, it also samplea the markers observed in its volume at the sensors arriving time. Then, it decides about the location based on the number of sensors which have released markers (the sensors with unfilled storages) and the number of sampled markers. Type-B FC has the ability of reading the level of sensors' storages and decides the abnormality location accordingly, without requiring to sample the markers.
	We propose optimum and sub-optimum decision rules and obtain the analytical probabilities of error, considering two cases of perfect and imperfect sensing regimes for the mentioned types of FCs. The results are also confirmed numerically.

	The rest of the paper is organized as follows. The system model is described in Section \ref{system_model}. Next, the problem description of abnormality detection is discussed in Section \ref{sec_detection_problem description}. The probabilities of false alarm and miss-detection are derived for the memoryless and aggregate sensors, which were also in some parts presented in our previous work \cite{khaloopour2020iwcit}. Then, the problem of abnormality localization using type-A and type-B FCs is investigated in this paper, in Section \ref{sec_localization_problem description}. The numerical results are provided in Section \ref{Sec: numerical}. Finally, Section \ref{Sec: conclusion} concludes the paper.
	

	\section{System Model} \label{system_model}
	
	We consider a cylindrical fluidic medium in a laminar flow condition, where the effective flow velocity is $v$. The medium is divided into some sensory regions, where at the end of each region, an FC is located to decide on the occurrence, and possibly the location of the abnormality in that region, e.g., $x_\text{FC}$ for the sensory region shown in Fig.~\ref{FCs_regions}. This process is similarly repeated for all regions. So, we only study the problem in the first region.
	Each sensory region is divided into some subregions with length of $vT$ (according to time-slots with duration $T$).
	In each sensory region, mobile sensors are used to detect the abnormality.
	For this purpose, at the beginning of the first subregion at the first region, \emph{i.e.}, $x_0$, there is an injection point, where $N_s$ fully charged sensors are released into the medium. Each sensor has an activation flag (denoted as $f_i\in\{0,1\}$ for $i=1,...,N_s$) to save the observation of abnormality. The sensors move to search the region to find the abnormality. As stated above, we consider a time-slotted system with slot duration of $T$.

	\subsection{Sensors Model} \label{seneors_model}
	The growing interest in the applications of sensory networks as well as the advances in the sensing devices is enabling the use of sensor networks to target challenging problems. While the sensors are mostly used in biological, medical, physiological and clinical applications, they are also used in industry, smart houses, mining, etc. Based on the kind of signal detected by sensors, different types of sensors may be considered such as electrical, magnetic, electrochemical, photonic, amperometry, and potentiometry with the capability of detecting various events, like pressure, stress, strain, ion, virus, and drug \cite{Rolfe2012, Unterweger2018}.
	In this paper, we propose a theoretical framework based on using mobile molecular nano-sensors that can detect an abnormality (depending on the application) in the fluidic medium. The theoretical model and the requirements of these sensors are explained in this subsection. For each requirement, we provide some possible candidates from the existing literature which might be useful for future practical implementations.
	
	\textbf{\emph{Movements:}}
	Our considered sensors are mobile nano-machines, moving in the fluidic medium thanks to diffusion and flow velocity.
	The importance of flow velocity over diffusion for the particles movements is explained by the Peclet number\footnote{The Peclet number is defined as
		$\textit{Pe}=\frac{v d_c}{D_p}$, where $d_c$ is the characteristic length of the channel and $D_p$ is the diffusion coefficient of the particles \cite{jamali2019channel}. $D_p$ is decreasing versus the particles mass and volume \cite{McGuiness2020}.}. For the sensors (nano-machines) which are relatively large particles, the Peclet number is much greater than 1 and the effect of flow velocity on their movements along the $\overrightarrow{x}$ direction is dominant over the diffusion \cite{jamali2019channel}. As a result, the sensors move with effective flow velocity $v$.
	Thus, it takes $K$ time-slots for the sensors to arrive the FC from their injection point, where $K=\lceil \frac{x_\text{FC}-x_0}{vT} \rceil$.

	In some applications like abnormality detection in pipelines and blood vessels, the physical access to each point of medium is not feasible. In these environments, the mobile sensors can move in the medium (such as bacterial chemotaxis \cite{okaie2014cooperative}), and reach at the vicinity of the abnormality point, enabling more precise and accurate sensing \cite{mosayebi2018early, Metin2017, Perfezou2012}.
	
	\textbf{\emph{Detection:}}
	In general, the abnormality may be a pollution leakage, pressure drop, harmful contaminates, obstacles, biomarkers and cancer cells. Based on the type of abnormality, the sensors are equipped by a sensing technology to detect the considered abnormality. The sensors' imperfect sensing is modeled probabilistically.
	If a sensor visits an actual abnormality, it is activated with the probability $\alpha$ at this point, and if no abnormality exists, it is activated with probability $\delta$ during each time-slot duration of $T$ (note that $\alpha \gg \delta$). Thus, each sensor has the probability of miss-detection equal to $(1-\alpha)$ and the probability of false alarm equal to $\delta$.
	
	In practice, the nanosensors respond faster, more accurate and precise, require lower power and voltage, and smaller materials \cite{Perfezou2012}.
	Different detection methods have been proposed in practical applications. For leakage detection in pipelines, the sensors may have microphones to detect the leaking sound\cite{Cai2020}. For detecting pressure drop in pipelines, they may have pressure sensing technology \cite{Zhu2017}. For detecting the contaminates in water pipes, or the biomarkers in blood vessels, they may have different receptors to detect the ligand molecules (contaminate, virus, bacteria, ion, and biomarker) \cite{Rolfe2012, Ghavami2017}. Then, a signal is generated by a transducer (changes in the current, voltage, ohmic properties) \cite{Perfezou2012}.
	In \cite{Ghavami2017}, this signal is chemical (releasing another type of molecules) and it is generated probabilistically, i.e., with probabilities $\alpha$ and $1-\alpha$ respectively in the presence and absence of an abnormality. In \cite{Yemini2020}, a sensor network is considered, where each sensor has probabilities of miss-detection $(1-\alpha)$, and false alarm $\delta$.
	
	\textbf{\emph{Storage:}}
	The sensors produce and store markers, to be released in order to activate the other sensors when detecting an abnormality.
	They have a limited marker production rate of $\beta$ and a maximum storage capacity, $M$. If the storage level $B_M$ is not full (i.e., $B_M < M$), it is charged linearly with the rate of $\beta$. The sensors' storages are fully charged at the injection time.
	We assume that the required time duration for charging an empty storage equals $K_s$ time-slots, where $K_s \geq K$ (\emph{i.e.,} $M=\beta K_sT$).
	
	This type of sensors is proposed in \cite{arjmandi2016ion} as a model for natural cells, according to their control mechanisms for marker generation and releasing. The markers are generated by chemical reactions until the concentration of markers inside the cell reaches a threshold of equilibrium state. This model is also used in \cite{Bafghi2018, khaloopour2019adaptive}.
	
	\textbf{\emph{Marker releasing:}}
	Since the number of sensors, $N_s$, is not very large and the activation of each sensor is not perfect (modeled by probabilities $(1-\alpha)$ and $\delta$), the sensors cooperate in detecting the abnormality by activating each other.
	It means that if a sensor senses an abnormality, it gets activated and releases all its stored markers ($M$ molecules) of a specific type at the beginning of the next time-slot to warn the others.
	For abnormality detection, these markers are used for cooperative sensors' activation. For abnormality localization, these markers can also be used by the FC to find the location of abnormality.
	
	There are various methods, proposed for releasing process in existing works. It is performed by opening the cell outlets in \cite{arjmandi2016ion}, where they can be opened or closed by applying a voltage or by reacting with ligand molecules. It also can be done by the reaction of transmitter gates with ligand molecules or applying a voltage (electrical field) on the outlet channels of the transmitter \cite{arjmandi2016ion}.
	Moreover, the stored molecules in a reservoir can be released by detecting a defects in the environment \cite{Unterweger2018}.
	
	\textbf{\emph{Sampling:}}
	An inactive sensor can be activated by sensing the markers observed in its volume at sampling times $t^s_j=jT+T_d, j=1,2,\cdots$, in all time-slots.
	The shift time $T_d$ is a design parameter which is added to let the released molecules diffuse in the environment and have a uniform concentration across the cross-section.
	We consider two types of sensors based on their decision strategy for activation by markers (memoryless and aggregate sensors), which will be explained later.
	
	The sensors can recognize the time-slots by their clock, which can be a negative transcription/ translation feedback loop for generating oscillations \cite{Fustin2013, Elowitz2000, Crnko2019}.
	The design and tuning the oscillating period in engineered bacteria are studied in \cite{ Novak2008, Stricker2008}. In the biological applications, the molecular clock can be regulated with heart rates \cite{Crnko2019}. The clock synchronization among nanomachines is also studied in \cite{Lin2015D, Lin2016A}.

	\textbf{\emph{Receiving}:}
	We assume that all sensors are received at the FC, for example magnetic nano-sensors can be absorbed by a magnet \cite{Xie2012, Cohen2013}. Their flags' values are read by the FC and then are reset to $f_i=0$, for all $i$.
	The FC waits for a duration in order to let all the sensors storages be completely charged. Then, it releases them to the next sensory region. The waiting time required to fill the storage of each sensor is used to measure its storage level at the arriving time, because of linearly molecule charging process with the rate of $\beta$. This information also can be used for the localization as will be described later.
	
	We consider the following mathematical models for the two noise sources.
	\\
	$1.$ \emph{Marker noise}:
	The molecules used as markers may exist in the environment from other sources. This is modeled as a Poisson noise with a fixed rate distributed in the region for all times \cite{ mosayebi2017cooperative}. The proof can be seen in \cite{jamali2019channel}. In the detection phase, these markers can be received by each sensor and cause the sensor to be (falsely) activated. In the localization phase, these markers can be received by the FC and cause localization error. We denote this noise rate as $\lambda$ and thus, the number of noise molecules is
	$
	n(t)\sim \text{Poisson}(\lambda).
	$
	\\
	$2.$ \emph{Sensor noise (imperfect sensing)}:
	The sensors can incorrectly be activated because of a wrong abnormality detection caused by imperfect sensing operation, within the subregion with time duration of $T$ with probability of $\delta$, as discussed above.
	
	In addition, we consider that the sensor may miss to detect an abnormality with probability $(1-\alpha)$.
	
	\subsection{Channel parameters}
	We study the laminar flow condition. More specifically, we assume that the effective flow velocity, $v$, is low enough and the equivalent diameter of the cylindrical channel, $d_e$, is small enough to have $d_e v < 2300\frac{\eta}{\rho}$, where $\rho$ and $\eta$ are respectively the density and the constant dynamic viscosity of the fluid.
	Therefore, the Reynolds number of the channel is less than $2300$ and the flow condition is laminar \cite{jamali2019channel}. The sensors are large nanomachines with respect to other particles. Thus, they have small diffusion coefficient, as mentioned before. Therefore, in this flow regime, the sensors move with the effective flow velocity $v$ along the $\overrightarrow{x}$ direction. 
	
	Also, we assume that $v a_c^2 \ll D_e \Delta x$, where $a_c$ is the radius of the pipe, $D_e$ is the effective diffusion coefficient, and $\Delta x$ is the distance of the releasing and the receiving points of the markers. Therefore, the dispersion factor of the molecules in the channel is much greater than $1$ \cite{jamali2019channel} and the concentration of the markers is uniform in the channel cross-sections at the receiving points.
	
	Now, we state the system models for the goals of abnormality detection and abnormality localization.
	
	
	\subsection{Abnormality Detection} \label{subsec:ab_det}
	
	For detecting an abnormality, the FC uses the number of active sensors, at the arriving point (i.e., $x_\text{FC}$). If this number is more than a threshold $\tau_1$, the FC warns the system that an abnormality is detected in its sensory region (see Fig. \ref{FCs_regions}). 	
	The sensors can be activated (the sensor's flag becomes ``1") in two ways: (i) the direct activation by sensing an abnormality in its moving path, (ii) cooperative activation by sensing the markers at sampling times $t^j_s$.
	The time-slot period $T$ is considered large enough in order to the sensor observations in different time-slots to be independent of each other \cite{jamali2019channel}. Note that the sensors that are activated by sensing the markers do not release further markers.
	Based on the activation strategy, we consider two types of sensors:\\
	$\bullet$ \textbf{Memoryless sensor}: it senses the environment in each sampling time $t^j_s$. If the number of observed markers in a sampling time is more than threshold $\tau_2$, it is activated. For example, the PH-meter \cite{khaloopour2019exp}, alcohol detectors \cite{farsad2013tabletop}, and magnetic detector for nano-particles \cite{Unterweger2018} are memoryless sensors.
	\\
	$\bullet$ \textbf{Aggregate sensor}: it senses the environment in each sampling time $t^j_s$ and aggregates the number of observed markers in these $K$ sampling times. If this value is more than threshold $\tau'_2$, it is activated. This type of sensors is considered in \cite{mosayebi2018early}.
	Note that the effects of both noises are appeared stronger for aggregate sensors compared to memoryless ones. This may result in high value of false alarm, \emph{i.e.}, $P_{e|H_0}$, for the aggregate sensors.
	Therefore, the threshold $\tau'_2$ must be set larger than $\tau_2$ to decrease $P_{e|H_0}$ for the aggregate sensors.

	\subsection{Abnormality Localization}
	Here, we describe the abnormality localization scenario. 
	We do not use cooperation among sensors for the localization, since this worsens the system performance.
	Remember that the sensors which are activated by markers (cooperative activation) do not release markers in the channel.
	As a result, the markers are only used by the FC.
	After detecting an abnormality in a region, to find its location, the FC uses the number of markers and the sensors storages levels.
	
	When a sensor is activated by the sensor noise or abnormality and releases $M$ stored molecules into the medium, it begins producing and storing molecules in its storage (\emph{i.e.,} $B^n_M$ for the $n$-th sensor).
	Therefore, if $n$-th sensor is activated by the noise or abnormality, its storage is not full ($B^n_M<M$), at the time of arriving the FC. In other word, the level of a sensor storage specifies its activation time ($t=t_s^K-\frac{B^n_M}{\beta}$).
	We consider two types of FCs based on their resolution of reading the sensors' storages level, as follows.
	\\
	$\bullet$ \textbf{Type-A FC}: it can read the flag values of the received sensors. Also, it can realize that each sensor's storage is full or not.
	\\
	$\bullet$ \textbf{Type-B FC}: it can read the flag values of the received sensors. Furthermore, it can also read each sensor's storage level.

	\begin{table}
		\centering
		\caption{Used Notations}
		\begin{tabular}{ c l}
			\hline Symbol & Description \\\hline
			$T$ & Time-slot duration\\
			$v$ & Flow velocity\\
			$x_\text{FC}$ & FC location \\
			$x_0$ & Sensors injection point \\
			$N_s$ & Total number of injected sensors \\
			$f_i$ & $i$-th sensor's activation flag\\
			$\alpha$ & Activation probability at the abnormality\\
			$\delta$ & Activation probability by imperfect sensing\\
			$t_j^j$ & $j$-th sampling time after the injection\\
			$T_d$ & Sampling shift time\\
			$\beta$ & Marker production rate\\
			$B^n_M$ & Storage level of the $n$-th sensor at the FC\\
			$M$ & Maximum markers capacity of a sensor's storage\\
			$K_s$ & Required time for charging an empty storage\\			
			$K$ & Total number of time-slots\\			
			$N_1$ & Number of sensors activated either by sensor noise or by\\
			& the abnormality\\
			$N_2$ & Number of sensors activated only by markers\\
			$N_T$ & Total number of activated sensors at the FC\\
			$\tau_1$ & Sensor's threshold on the number of sampled markers\\
			$\tau_2$ & FC's threshold on the number of activated sensors\\
			$V_s$ & Sensors' volume for marker sampling\\
			$V_\text{FC}$ & FC's volume for marker sampling\\
			$D$ & Markers diffusion coefficient in the fluid\\
			$a_c$ & Radius of the channel\\
			$X^*$ & Abnormality location\\
			$R_i$ & Number of sensors activated either by sensor noise or by\\
			& the abnormality in $i$-th time-slot\\
			$p_i$ & Sensors' activation probability either by sensor noise or by\\
			&  the abnormality  in $i$-th time-slot\\
			$\gamma_i$ & FC's thresholds for localization\\
			\hline
		\end{tabular}
		\label{table_state_1ISI}
		\vspace{-0.5em}
	\end{table}
	
	\section{Abnormality Detection}\label{sec_detection_problem description}
	
	In this section, we focus on abnormality detection. The FC detects an abnormality based on the states of absorbed sensors.
	Total number of active sensors at the FC is $N_T = N_1+N_2$,
	where $N_1$ is the total number of sensors activated either by the abnormality (direct activation) or by imperfect sensing, and $N_2$ is the number of sensors activated only by markers (cooperative activation). Note that $N_1$, $N_2$ and $N_T$ are random variables (RVs), and their realizations are shown by $n_1$, $n_2$ and $n_T$, respectively. The FC receives all sensors and reads their activation flags to find $N_T$. If $N_T \geq \tau_1$, it declares an abnormality. Our approach is to analyze the system performance based on error probability metrics.
	
	We denote the number of sensors activated either by the imperfect sensing (sensor noise) or by the abnormality (direct activation) in $i$-th time-slot by the RV $R_i$. We also define the random marker source vector as $R=[R_1, R_2, \ldots, R_K]$. We know that $N_1=\sum _{i=1}^{K}R_i$. Each activated sensor releases the markers at the end of the activation slot. The other sensors, which are activated neither by sensor noise nor by direct activation, may be activated by markers. Thus, the source vector $R$ shows all the marker sources in all slots and has $K$ elements $R_1,R_2,\cdots,R_K$.
	The distribution of these elements is not obtained in our previous work \cite{khaloopour2020iwcit}.
	If we define the $R_{K+1}=N_s-\sum_{i=1}^{K} R_i$ as the number of sensors that are activated neither by abnormality nor by the sensor noise, the $R_i$s, $i=1,\cdots, K+1$ have a multinomial distribution with corresponding probabilities $p_1,p_2,\cdots,p_{K+1}$, which will be obtained in the next subsection.
	
	We have a hypothesis testing problem (in each FC) for the existence of abnormality with null hypothesis $ {H}_0$ and alternative hypothesis $ {H}_1$. The probability of detection error~is
	\begin{align}
	{P}^\text{D}_e = {P}( {H}_0){P}^\text{D}_{e| {H}_0}+{P}( {H}_1){P}^\text{D}_{e| {H}_1} , \label{Eq: Pe}
	\end{align}
	where ${P}_{e| {H}_0}$ and ${P}_{e| {H}_1}$ are known as the probabilities of false alarm and miss-detection. To obtain these probabilities, we need the distribution of number of active sensors received by an FC, for two hypothesis $ {H}_0$ and $ {H}_1$.
	Now, we derive the error probability for the memoryless and aggregate sensors.
	
	\subsection{Memoryless Sensors} \label{Subsec: Memoryless sensors}
	
	As mentioned before, there are two noise sources: marker noise and imperfect sensing.
	The imperfect sensing increases the probability of a sensor being activated. The sensors, activated by the imperfect sensing, release markers at the beginning of following time-slot (for cooperative activation). As a result, other sensors can mistakenly be activated by receiving the markers.
	In the following, we derive the probabilities of false alarm and miss-detection in the presence of imperfect sensing and marker noises.
	
	\subsubsection{\textbf{False alarm}} \label{Subsubsec: FA_2}
	Consider the hypothesis $ {H}_0$. Remind that due to imperfect sensing a sensor can be activated with probability $\delta$ in each time-slot when passing the distance of $d = vT$ along the $x$-axis. Here, no abnormality exists and thus the probabilities $p_1,p_2,\cdots,p_{K+1}$ are as follows.
	\begin{equation}
	p_i= \left\{
	\begin{array}{rl}
	\delta(1-\delta)^{i-1} ,&   i=1,\cdots, K,
	\\
	(1-\delta)^{K}\text{ }\text{ }\text{ },&  i= K+1
	\end{array} \right.. \nonumber
	\end{equation}
	If we define $r=[r_1, r_2, \ldots, r_K]$, where $\sum_{i=1}^K r_i \leq N_s$, as a realization of $R$, the probability mass function (pmf) of $R$ is
	\begin{align}
	p_R(r) =&\text{Pr}\{R=r\}=\frac{N_s!}{r_1!\cdots r_{K+1}!}p_1^{r_1}\cdots p_{K+1}^{r_{K+1}}.
	\nonumber
	\end{align}

	If $\{R=r\}$ occurs, the number of sensors activated only by markers (cooperative activation) is distributed as
	$ N_2 \sim \text{Binomial}(N_s- N_1,P_{\text{active}|R})$, where $P_{\text{active}|R}$ is the probability that a sensor is activated only by markers before it reaches the FC and is obtained as
	\begin{align}
	P_{\text{active}|R} =1-\prod_{i=1}^{K} (1-{P}_{a,i|R}), \label{P_active|R}
	\end{align}
	where $P_{a,i|R}$ is the probability of cooperative activation in $i$-th slot. Thus, we have $P_{a,i|R}= \text{Prob}[Y(t^i_s) \geq \tau_2]$, where $Y(t^i_s)$ is the number of observed markers by a sensor in $i$-th sampling time. Remind that the markers sensed by a sensor in each slot are released in previous slots. Thus, we have
	\begin{align}
	Y(t^i_s) \sim \text{Poisson} (\sum_{j=1}^{i}r_jM\mu_{ji}+\lambda ), \label{Yi_ from R}
	\end{align}
	where $\mu_{ji}$ is the probability of a marker released at $t=jT$ to be observed in the sensor's volume $V_\text{s}$ at $i$-th sampling time ($t^i_s$). The $\mu_{ji}$ generally depends on the location of sensors at times $jT$ and $t^i_s=iT+T_d$.
	Assuming that the sensors move with velocity $v$ as the flow (the laminar flow condition and flow dominant regime) and also assuming that the concentration of molecules is uniform in a cross-section of the cylindrical channel with radius $a_c$, the $\mu_{ji}$ is obtained as \cite{jamali2019channel}:
	\begin{align}
	\mu_{ji} = \frac{V_\text{s}}{4\pi a_c^2} \times \frac{1}{\sqrt{4\pi D ((i-j)T+T_d)})}, \text{ for }i\geq j, \label{MU equation}
	\end{align}
	where $D$ is the diffusion coefficient of the markers in the fluid.
	
	The number of released markers in \eqref{Yi_ from R} is large enough that we can use Gaussian approximation for Poisson distribution \cite{jamali2019channel}, which results in
	\begin{align}
	{P}_{a,i|R=r}= \text{Q} (\frac{\tau_2-\sum_{j=1}^{i}r_jM\mu_{ji}-\lambda}{\sqrt{\sum_{j=1}^{i}r_jM\mu_{ji}+\lambda}} ). \label{P_{a|R_H0_A}}
	\end{align}
	A false alarm occurs when $N_T=N_1+N_2 \geq \tau_1$ (\emph{i.e.}, $N_1 \geq \tau_1$ or $N_2 \geq \tau_1-N_1$). Therefore, the probability of false alarm is
	\begin{align}
	P^\text{D}_{e| {H}_0} & = \sum_{r} p_R(r){P}^\text{D}_{e| {H}_0,R}
	=\sum_{r:n_1 \geq \tau_1} p_R(r) +\sum_{R:n_1 < \tau_1} p_R(r) \times
	\nonumber
	\\ &\sum_{j=\tau_1-n_1}^{N_s-n_1}{{N_s-n_1}\choose {j}}P_{\text{active}|R=r}^j(1-P_{\text{active}|R=r})^{N_s-n_1-j}.
	\label{P_e|H0  sensor noise}
	\end{align}

	\subsubsection{\textbf{Miss-detection}} \label{Subsubsec: MD_2}
	
	Now, we assume that there is an abnormality in the region and  the sensors may be activated by the abnormality (both direct and cooperative activation), sensor noise, and marker noise.
	We assume that the abnormality occurs at point $X^*$, which is an RV.
	We denote the number of slots it takes for the sensors to reach the abnormality by $J^*$. Because of the flow dominant assumption, the sensors move with velocity $v$ along the channel and we have $J^*= \lceil \frac{X^*-x_0}{v} \rceil$. Thus $J^*$ is a uniform RV as $J^* \sim \text{Unif}[1:K]$.
	
	If $r=[r_1,\ldots, r_{J^*}, \ldots,r_K]$ be the realization vector of $R$, then each component $r_{i}$, for $i \neq J^*$, is a result of sensor noise while $r_{J^*}$ is due to both the noise and the abnormality independently.
	Therefore, the corresponding probabilities are
	\begin{equation}
	{p}_i= \left\{
	\begin{array}{rl}
	\delta(1-\delta)^{i-1} \text{ }\text{ }\text{ }\text{ }\text{ }\text{ }\text{ }\text{ }\text{ }\text{ }\text{ }\text{ }\text{ },&   i=1,\cdots, J^*-1
	\\
	(\alpha+\delta)(1-\delta)^{J^*-1}\text{ }\text{ }\text{ }\text{ },&   i= J^*
	\\
	\delta(1-\alpha-\delta)(1-\delta)^{i-2},&  i= J^*+1,\cdots, K
	\\
	(1-\alpha-\delta)(1-\delta)^{K-1},&  i= K+1
	\end{array} \right. ,\nonumber
	\end{equation}
	{which are not obtained in our previous work \cite{khaloopour2020iwcit}.}
	The conditional pmf of vector $R$ is
	\begin{align}
	p_R(r|J^*) =&\text{Pr}\{R=r|J^*\}=\frac{N_s!}{r_1!\cdots r_{K+1}!}{p}_1^{r_1}\cdots {p}_{K+1}^{r_{K+1}}.
	\label{P_R, with sensornoise_A, H1}
	\end{align}

	If $\{R=r\}$ occurs, the number of molecules that a sensor senses at $i$-th sampling time ($t_s^i$) is obtained as \eqref{Yi_ from R}.
	Therefore, the probability of a sensor to be activated before it reaches the FC ($P_{\text{active}|R}$) is obtained from \eqref{P_active|R}. Thus, we have
	\begin{align}
	&{P}^\text{D}_{e| {H}_1,J^*}  = \sum_{R} p_R(r|J^*){P}^\text{D}_{e| {H}_1,R=r}
	=\sum_{r:n_1 =0}^{n_1 < \tau_1} p_R(r|J^*) \times \nonumber \\
	& \sum_{j=0}^{\tau_1-n_1-1} {{N_s-n_1}\choose {j}}P_{\text{active}|R=r}^j (1-P_{\text{active}|R=r})^{N_s-n_1-j}.
	\label{P_e|H1,x*,  sensor noise A}
	\end{align}
	We have $J^*\sim \text{Unif}[1:K]$. Thus, we obtain ${P}^\text{D}_{e| {H}_1}$ as
	\begin{align}
	&{P}^\text{D}_{e| {H}_1}  = \frac{1}{K}\sum_{j^*=1}^{K}  {P}^\text{D}_{e| {H}_1,J^*}.
	\label{P_e|H1, sensor noise A}
	\end{align}
	
	\subsection{Aggregate Sensors} \label{Subsec: Aggregate sensors}
	In this case, the sensors are activated by the summation of the number of observed markers in all previous time-slots. Thus, we only need to find the  summation in the last time-slot, before a sensor reaches the FC.
	
	\subsubsection{\textbf{False alarm}}
	Similar to Subsection \ref{Subsubsec: FA_2}, we assume that the marker source vector is $R=[R_1, R_2, \ldots, R_K]$ with the realization vector of $r=[r_1, r_2, \ldots, r_K]$. Each sensor obtains the summation of markers received in all sampling times ($t^i_s=iT+T_d, \text{ }i=1,\ldots,K$) as,
	\begin{align}
	Y(t^i_K) \sim \text{Poisson} ( \sum_{i=1}^{K} \sum_{j=1}^{i} r_jM\mu_{ji}+K\lambda ). \label{poiss_2}
	\end{align}
	Using Gaussian approximation for \eqref{poiss_2} results in
	\begin{align}
	Y(t^i_K) \sim \mathcal{N}  ( \sum_{i=1}^{K} \sum_{j=1}^{i} r_j M\mu_{ji}+K\lambda,\sqrt{\sum_{i=1}^{K} \sum_{j=1}^{i} r_j M\mu_{ji}+K\lambda}). \nonumber
	\end{align}
	The probability of an inactive sensor to be activated only by markers, passing a region, is
	\begin{align}
	P_{\text{active}|R}=P_{{a},K|R},
	\label{1_P_active|R}
	\end{align}
	where $P_{{a},K|R}$ is obtained as follows.
	\begin{align}
	P_{{a},K|R=r}=\text{Q}(\frac{\tau'_2- \sum_{i=1}^{K} \sum_{j=1}^{i} r_j M\mu_{ji}-K\lambda}{( \sum_{i=1}^{K} \sum_{j=1}^{i} r_j M\mu_{ji}+K\lambda)^{0.5}}). \nonumber
	\end{align}
	The ${P}^\text{D}_{e| {H}_0}$ is obtained by substituting \eqref{1_P_active|R} in \eqref{P_e|H0  sensor noise}.

	\subsubsection{\textbf{Miss-detection}}
	
	Similar to Section \ref{Subsubsec: MD_2}, in this subsection, we assume that $\alpha \gg \delta$, and the marker source vector is $R=[R_1,R_2,\ldots, R_{J^*}, \ldots,R_K]$ with realization vector of $r=[r_1,r_2,\ldots, r_{J^*}, \ldots,r_K]$, where $r_{J^*}$ is resulted from the sensor noise and abnormality with independent effects. Thus, the probability of this vector is given in \eqref{P_R, with sensornoise_A, H1}.
	${P}^\text{D}_{e| {H}_1,J^*}$ is obtained by substituting \eqref{1_P_active|R} in  \eqref{P_e|H1,x*,  sensor noise A}. Then, the probability of miss-detection ${P}^\text{D}_{e| {H}_1}$ is obtained by substituting ${P}^\text{D}_{e| {H}_1,J^*}$ in \eqref{P_e|H1, sensor noise A}.

	\section{Abnormality Localization} \label{sec_localization_problem description}
	
	In this section, we assume that an abnormality has occurred in a region and has been detected by the related FC. Now, the FC wishes to find the location of the abnormality. We need more capable FCs for the localization. We assume that each FC can sample the markers at $t= t^s_K$. Moreover, the FC knows the states of sensors' storages with some resolution.
	Based on the number of received markers and the states of sensors' storages, the FC decides about the location of abnormality.
	
	We study two types of FCs based on their knowledge about the received sensors' storage level, as follows.
	\\
	\textbf{1. Type-A FC:}
	In addition to reading the flags value of the received sensors, it realizes whether each sensor's storage is full or not. The sensors, whose storages are not full, have been activated either by sensor noise or by the abnormality (\emph{i.e.}, $\sum R$) and are the sources of markers. Note that a type-A FC cannot find the abnormality location without using markers. Because, though it knows the total number of activated sensors, it does not know in which subregion each sensor is activated.
	\\
	\textbf{2. Type-B FC:}
	In addition to the flags value of the received sensors, it reads the level of each sensor's storage.
	As said before, if a sensor is activated by abnormality or by imperfect sensing, it releases all its stored molecules into the medium. Then, its storage is linearly charged with a limited rate of $\beta$, and we assume $M>\beta K T$ (which guarantees that if a sensor releases markers into the channel, its storage is not full, when arriving at the FC).
	Since the sensors move in the medium with flow velocity of $v$, they travel a distance of $vT$ through the channel in each time-slot. Hereby, the storage level specifies the activation time-slot of each sensor.
	We assume that the abnormality occurs uniformly along the region.
	
\begin{figure}
		\centering
		\includegraphics[trim={4cm 8.7cm 1cm 5.5cm},clip,
		scale=0.3]{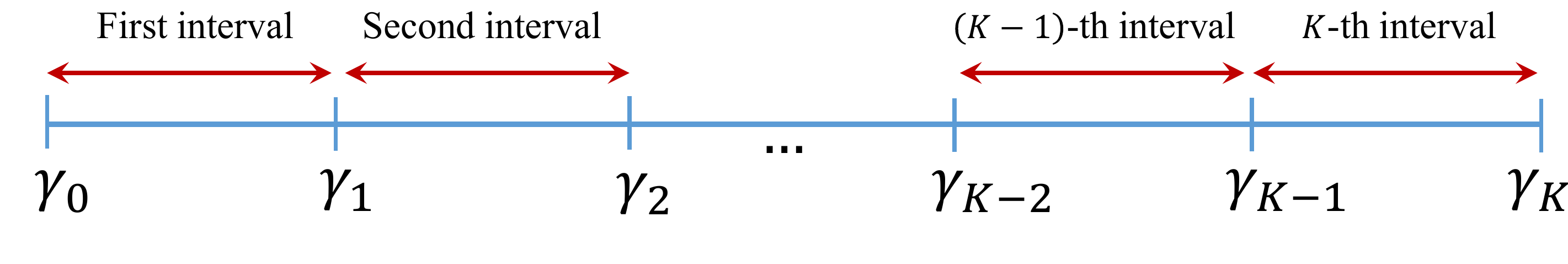}
		\vspace{-1em}
		\caption{The thresholds $\gamma_0, \dots , \gamma_{K}$ and the detected subregions.}
		\label{fig: all thr}
		\vspace{0em}
	\end{figure}

	For the localization, we are required to find the probability of receiving a marker by the FC.
	According to \eqref{MU equation}, the probability of receiving a marker by a receiver (sensor) is proportional to its volume\footnote{We assume that the receiving volume of FC is small enough that markers concentration is uniform there.}. Thus, if the volume of FC for receiving markers is $V_\text{FC}$, then the marker noise rate observed by the FC will be $\lambda'=\frac{V_\text{FC}}{V_\text{s}}\lambda$\footnote{Note that the parameter $\lambda$ is the marker noise rate observed by each sensor with receiving volume of $V_\text{s}$.}. Also the concentration of markers released by a sensor in $i$-th time-slot at the FC in $K$-th time-slot is $M\mu'_{iK}=\frac{V_\text{FC}}{V_\text{s}}M\mu_{iK}$\footnote{If we assume $V_\text{FC}=V_\text{s}$, then we have $\mu'_{iK}=\mu_{iK}$ and $\lambda'=\lambda$.}.
	Therefore, if a realization of 
	$R$ be $r=[r_1, r_2, \cdots, r_K]$, the number of received markers by the FC is
	\begin{align}
	Z_{\text{FC}} \sim \text{Poisson}(\sum_{i=1}^{K} r_i M\mu'_{iK}+\lambda'). \label{Poiss_markers at FC_loc}
	\end{align}
	Based on the number of received markers $Z_\text{FC}$, the FC detects the subregion where the abnormality occurs. There is $K$ slots. Because the sensors movements are considered in flow dominant regime with flow velocity of $v$, the sensors move equal distances in each time-slot, which means that we have $K$ subregions. Thus, the FC faces $K$ hypotheses $\tilde H_1,\cdots, \tilde H_K $, where $\tilde H_i$ means $J^*=i$. It uses $K+1$ thresholds $\gamma_0=0,\gamma_1, \dots, \gamma_{K-1}, \gamma_K=\infty$ to find the abnormality subregion (see Fig. \ref{fig: all thr}). In other word, the detected subregion is as follows.
	\begin{equation}
	\hat{J}^*= \left\{
	\begin{array}{rl}
	1,  &  \gamma_0 \leq Z_\text{FC} < \gamma_1\\
	&     \vdots
	\\
	K, &  \gamma_{K-1} \leq Z_\text{FC} < \gamma_{K}
	\end{array} \right..
	\label{hypothesis_testing_prob1}
	\end{equation}
	An error occurs when the detected subregion by the FC is not true ($\hat{J}^* \neq J^*$).
	Thus, the probability of localization error is
	\begin{align}
	P_{e|R}^{\text{L}}&=\sum_{j=1}^{K}P(J^*=j)P_{e|R,J^*=j}^{\text{L}} \nonumber\\
	&=\frac{1}{K}\big(\sum_{j=1}^{K+1}\overline{P(\gamma_{j-1}\leq Z_\text{FC} < \gamma_{j})}\big),  \label{total_1}
	\end{align}
	where $\overline{P(\cdot)}=1-{P}(\cdot)$. We know that the components $R_i$s of the marker source vector $R$ are random. It means that the number of sensors releasing markers into the medium is random. Also these sensors may be activated in random slots (or subregions). Therefore, the parameter of Poisson distribution in \eqref{Poiss_markers at FC_loc} is random and the term of $\overline{P(\gamma_{j-1}\leq Z_\text{FC} < \gamma_{j})}$ in \eqref{total_1} may have a complex form. To simplify this term, we first consider the case of perfect sensing where $\delta=0$. Then we consider the general case of imperfect sensing where $\delta >0$.	
	
	
	\subsection{Perfect sensing: $\delta=0$}
	\label{Perfect sensing _type B}
	In this case, no sensor is activated because of imperfect sensing. So, the random marker source vector is $R=[R_1,\cdots,R_K]$, where $R_i=0, \forall i \neq J^*$ and $R_{J^*}>0$ with realization vector $r=[r_1,\cdots,r_K]$. Thus, we have
	\begin{align}
	Z_\text{FC} \sim \text{Poisson}( r_{J^*} M\mu'_{J^*K}+\lambda'). \label{poiss_marker_loc}
	\end{align}
	From \eqref{total_1} the $P_{e}^{\text{L}}$ can be written as
	\begin{align}
	P_{e}^{\text{L}}=\frac{1}{K}\big(&\sum_{r_{J^*}=1}^{N_s} p_{R_{J^*}}(r_{J^*}) \times \nonumber \\& \sum_{j=1}^{K+1} \overline{P(\gamma_{j-1}\leq Z_{\text{FC}|R_{J^*}=r_{J^*},J^*=j} < \gamma_{j})}\big),  \label{total_2}
	\end{align}
	where $\overline{P(\gamma_{j-1}\leq Z_{\text{FC}|R_{J^*}=r_{J^*},J^*=j} < \gamma_{j})}$ can be obtained from \eqref{poiss_marker_loc}. If we use Gaussian approximation for \eqref{poiss_marker_loc}, we have
	\begin{align}
	&\overline{P(\gamma_{j-1} \leq Z_{\text{FC}|R_{J^*}=r_{J^*},J^*=j} < \gamma_{j})}=\label{complement_prob_loc} \\
	&1-Q(\frac{\gamma_{j-1}-r_{J^*}M\mu'_{jK}-\lambda'}{\sqrt{r_{J^*}M\mu'_{jK}+\lambda'}})+Q(\frac{\gamma_{j}-r_{J^*}M\mu'_{jK}-\lambda'}{\sqrt{r_{J^*}M\mu'_{jK}+\lambda'}}).
	\nonumber
	\end{align}
	\\
	Now, we study the performance of type-A and type-B FCs.
	
	\subsubsection{\textbf{Type-A FC}}
	For this type, we want to obtain the optimum thresholds which minimize the $P_{e}^{\text{L}}$ in \eqref{total_2}. As mentioned, this FC knows whether the storages of received sensors are full or not (\emph{i.e.}, $\sum R$). The sensors with unfilled storages are those who released markers at the abnormality point. Since $\sum R=R_{J^*}$, this type of FC knows the number of sensors which are activated at the $J^*$-th time-slot ($R_{J^*}$) and the marker source vector is $R=[0,\cdots,0,R_{J^*},0,\cdots,0]$. Assuming $r=[0,\cdots,0,r_{J^*},0,\cdots,0]$ is its realization vector, we can minimize $P_{e|R}^{\text{L}}$ in \eqref{total_1} instead of $P_{e}^{\text{L}}$.
	We can find optimum thresholds $\gamma^{*}_j$s by solving the following equation.
	\begin{align}
	\frac{d}{d\gamma_{j}}&P_{e|R=r}^{\text{L}}= \label{gama_eq}\\
	&\frac{1}{K}\Big(\frac{-1}{\sqrt{r_{J^*}M\mu'_{(j+1)K}+\lambda'}}\frac{d}{dx}Q(x)\Big|_{\frac{\gamma^{*}_{j}-r_{J^*}M\mu'_{({j+1})K}-\lambda'}{\sqrt{r_{J^*}M\mu'_{({j+1})K}+\lambda'}}} \nonumber \\
	&+\frac{1}{\sqrt{r_{J^*}M\mu'_{jK}+\lambda'}}\frac{d}{dx}Q(x)\Big|_{\frac{\gamma^{*}_{j}-r_{J^*}M\mu'_{jK}-\lambda'}{\sqrt{r_{J^*}M\mu'_{jK}+\lambda'}}}\Big)=0,
	\nonumber
	\end{align}
	where $\frac{d}{dx}Q(x)= \frac{-1}{\sqrt{2\pi}}\exp(-\frac{x^2}{2})$. Since $r_{J^*}$ is known at the FC, the optimal thresholds are obtained:
	\begin{align}
	\gamma^{*}_{j}=
	&\Big((r_{J^*}M\mu'_{jK}+\lambda)(r_{J^*}M\mu'_{(j+1)K}+\lambda) \nonumber \\&\big( \frac{\ln(\frac{r_{J^*}M\mu'_{(j+1)K}+\lambda}{r_{J^*}M\mu'_{jK}+\lambda})}{r_{J^*}M(\mu'_{(j+1)K}-\mu'_{jK})}+1 \big)\Big)^{0.5}.  \label{gama_opt}
	\end{align}
	The probability of error $P_{e}^{\text{L},A}$ is obtained by substituting the above thresholds in \eqref{total_2}.
	
	\subsubsection{\textbf{Type-B FC}}
	This type of FCs knows the storage level of the received sensors. As mentioned, the sensors produce markers linearly with limited rate of $\beta$ up to storage capacity. So, by knowing the storage level $B^n_M$ of the $n$-th sensor, the FC can find its activation time as $t=t_s^K-\frac{B^n_M}{\beta}$. In case of perfect sensing ($\delta=0$), we have $r_i=0, \forall i \neq J^*$. Therefore, all sensors which released markers in a region are activated at the abnormality point ($J^*$-th time-slot) and have equal storages levels\footnote{Note that the localization is performed if the abnormality is detected by the FC. Thus, at least one sensor has been activated at the abnormality location.}. So the FC can find the location of the abnormality as $ \lceil v(t_s^K-\frac{B^n_M}{\beta})-1 ,v(t_s^K-\frac{B^n_M}{\beta})\rceil$ (in $\lceil  \frac{t_s^K}{T}-\frac{B^n_M}{\beta T} \rceil $-th time-slot), with noo error (\emph{i.e.}, $P_e^\text{loc, B}=0$).

	
	\subsection{Imperfect sensing: $\delta>0$}
	
	The problem of localization in imperfect sensing regime is more complex than perfect sensing regime. Because random number of sensors can be activated at all time-slots. That is the markers are released not only at the $J^*$-th slot but also at other sensor activation slots, which makes the localization difficult. Now, we study the performance of type-A and type-B FCs.
	
	\subsubsection{\textbf{Type-A FC}}
	The FC knows the total number of sensors that are activated either by the abnormality or by the sensor noise, which released markers in the medium. The probability of localization error for this FC is
	\begin{align}
	P_{e}^{\text{L},A,\delta}= \frac{1}{K+1}\Big(\sum_{j=1}^{K}\sum_{R}^{}p_R(r|J^*=j)P_{e|R,J^*=j}^{\text{L},A,\delta}\Big), \label{Pe_new A,delta}
	\end{align}
	where by Gaussian approximation, we have
	\begin{align}
	P_{e|R,J^*=j}^{\text{L},A,\delta}&= 1-Q(\frac{\gamma_{j-1}-\sum_{i=1}^{K} r_iM\mu'_{iK}-\lambda'}{\sqrt{\sum_{i=1}^{K} r_iM\mu'_{iK}+\lambda'}})\nonumber \\
	\text{ }&+Q(\frac{\gamma_{j}-\sum_{i=1}^{K} r_iM\mu'_{iK}-\lambda'}{\sqrt{\sum_{i=1}^{K} r_iM\mu'_{iK}+\lambda'}}),
	\label{Pe_loc_B_delta}
	\end{align}
	where $\gamma_i$s are the thresholds used in \eqref{hypothesis_testing_prob1} and will be determined in the following.
	We define $N_3$ as the number of sensors which are activated by imperfect sensing  in all slots except $J^*$-th slot. In other words, $N_3=\sum_{i=1,\cdots,K, i\neq J^*} R_i$ and $R_{J^*}>0$.
	If $p_{N_3}(n_3)$ shows the pmf of RV $N_3$, the probability of error is
	\begin{align}
	P_{e}^{\text{L},A,\delta}=p_{N_3}(0)P_{e|N_3=0}^{\text{L},A,\delta}+(\sum_{n_3=1}^{N_s}p_{N_3}(n_3)) P_{e|N_3>0}^{\text{L},A,\delta}.
	\nonumber
	\end{align}
	In this case ($\delta>0$), since the markers are released from different locations (abnormality location and sensor noise locations), there are many possible realizations for $R$. For finding the optimal thresholds, we need to find the derivative of \eqref{Pe_new A,delta} and solve $dP_{e}^{\text{L},A,\delta}/dt=0$. Due to large number of possible realizations for $R$, finding the optimal thresholds $\gamma_j$s in a closed form is not feasible.
	On the other hand, the type-A FC only knows the total number of sensors activated either by the abnormality (at the actual subregion) or by the sensor noise (at other normal subregions), \emph{i.e.}, $\sum_{i=1}^{K}r_i$. So it cannot distinguish between the elements $r_1,r_2,\cdots,r_K$.
	Thus, we propose a feasible method to find suboptimum thresholds. We approximately assume that all of received sensors with unfilled storages are activated at the abnormality point ($r^\text{New}_{J^*}=\sum_{i=1}^{K} r_i$), and obtain suboptimum thresholds  by substituting $r^\text{New}_{J^*}$ instead of $r_{J^*}$ in \eqref{gama_opt}.
	If $N_3=0$, no sensor is activated by imperfect sensing, and they are only activated at $J^*$-th time-slot. Thus, our above assumption is true ($r^\text{New}_{J^*}=r_{J^*}$) and $P_{e|N_3=0}^{\text{L},A,\delta}$ is the same as $P_{e}^{\text{L},A}$. Using $\sum_{n_3=1}^{N_s}p_{N_3}(n_3)=1-p_{N_3}(0)$ we have
	\begin{align}
	P_{e}^{\text{L},A,\delta}=p_{N_3}(0)P_{e}^{\text{L},A}+(1-p_{N_3}(0))P_{e|N_3>0}^{\text{L},A,\delta},
	\nonumber
	\end{align}
	where $p_{N_3}(0)$ indicates the probability of $r=[0,\cdots,0, r_{J^*}> 0,0,\cdots,0]$. Thus, from \eqref{P_R, with sensornoise_A, H1}, we have
	\begin{align}
	p_{N_3}(0)
	=&(1-\delta)^{N_s(K-1)-r_{J^*}(K-J^*)}.
	\nonumber
	\end{align}
	Using $P_{e|N_3>0}^{\text{L},A,\delta}<1$, an upper bound on $P_{e}^{\text{L},A,\delta}$ is obtained as
	\begin{align}
	P_{e}^{\text{L},A,\delta} \leq 1-p_{N_3}(0)(1-P_{e}^{\text{L},A}).
	\label{Pe_loc_delta not 0}
	\end{align}
	
	The performance of type-A FC depends on $M$. Now, we consider the most general case, for which the probability of localization error in imperfect sensing regime is derived in \eqref{Pe_new A,delta}. We know that the probability of localization error decreases versus the number of markers. 	
	The asymptotic behavior of $P_e^{\text{L}, A, \delta}$ when $M\rightarrow \infty$ is described in the following lemma, whose proof is provided in Appendix \ref{proof of lemma 1 and 2}.
	\begin{lemma} \label{Lemma_1}
		In the perfect sensing regime, the probability of localization error for type-A FC is vanishing when $M\rightarrow \infty$, while in the imperfect sensing regime, it is non-vanishing when $M\rightarrow \infty$.
	\end{lemma}
	
	\subsubsection{\textbf{Type-B FC}}
	
	As mentioned, this type of FCs knows the storage level of the received sensors ($B^n_M$). The sensors that are activated either by the abnormality or by imperfect sensing release markers in the medium. The storage of these sensors is not full at the FC, since we have assumed that the required time for charging each sensor's storage equals $K_sT$, where $K_s>K$.
	By reading the $n$-th sensor's storage ${B^n_M}$, the FC can find its activation time as $t=t_s^K-\frac{B^n_M}{\beta}$ for $n=1,2,\cdots$, and find the marker source vector $R$. Assume that $r=[r_1, \cdots, r_K]$ is observed. The FC makes decision about the abnormality location by using the maximum a posteriori (MAP) rule as:
	$
	\underset{m=1,\cdots, K}{\operatorname{argmax}} \text{ } P({J}^*=m|R). \nonumber
	$
	We have assumed that the abnormality is distributed uniformly in a region. Thus,  the MAP rule will be reduced to maximum likelihood (ML) rule as
	$	
	\underset{m=1,\cdots, K}{\operatorname{argmax}} \text{ } p_R(r|{J}^*=m). \nonumber
	$
	Using \eqref{P_R, with sensornoise_A, H1}, the ML rule will~be
	\begin{align}
	\hat{J}^*=\underset{m=1,\cdots, K}{\operatorname{argmax}} \text{ } (1+\frac{\alpha}{\delta})^{r_m}(1-\frac{\alpha}{1-\delta})^{-\sum_{i=1}^{m}r_i}. \label{ML_total_type_B}
	\end{align}
	The decision rule in \eqref{ML_total_type_B} can be simplified for sufficient small $\delta$s as described in the following theorem. The proof is provided in Appendix \ref{Appx: ML_type B}.
	\begin{theorem}
		\label{Theorem_1}
		If we have
		$(1+\frac{\alpha}{\delta})(1-\frac{\alpha}{1-\delta})^{N_s-1}>1$,
		the ML decision rule in \eqref{ML_total_type_B} is reduced to
		$\hat{J}^*=\max \underset{m=1,\cdots, K}{\operatorname{argmax}} \text{ }r_m. \nonumber$
	\end{theorem}

	Now, we find the probability of error for type-B FC in the imperfect sensing case. An error occurs when $\hat{J}^* \neq {J}^*$. So the probability of error is
	\begin{align}
	P_{e}^{\text{L},B} &=\sum_{j=1}^{K} P(J^*=j)P_{e|J^*=j}^{\text{L},B}   \nonumber\\
	&=\frac{1}{K}\sum_{j=1}^{K} \sum_{R|j} P(\hat{J}^* \neq J^*|R,J^*=j). \nonumber
	\end{align}

	\section{Numerical Results} \label{Sec: numerical}
	
	In this section, we provide the numerical results for the proposed system for the parameters as  $v=0.02$ m/s, $a_c =0.05$ m, $V_s=V_\text{FC}=10^{-9}\text{ m}^3$, $\alpha=0.3$, $x_0=0$, $x_\text{FC}=600$ m, $N_s=20$, $M=10^7$ mole, $\lambda=\lambda' =5$ mole, $\delta =0.002$, $\tau_2=20$, $T=3000$ s, $T_d=2700$ s, $D=10^{-6} \text { m}^3\text{/s}$, $P(H_0)=0.1$.
	
\begin{figure}
		\vspace{-1em}
		\centering
		\includegraphics[height=138pt]{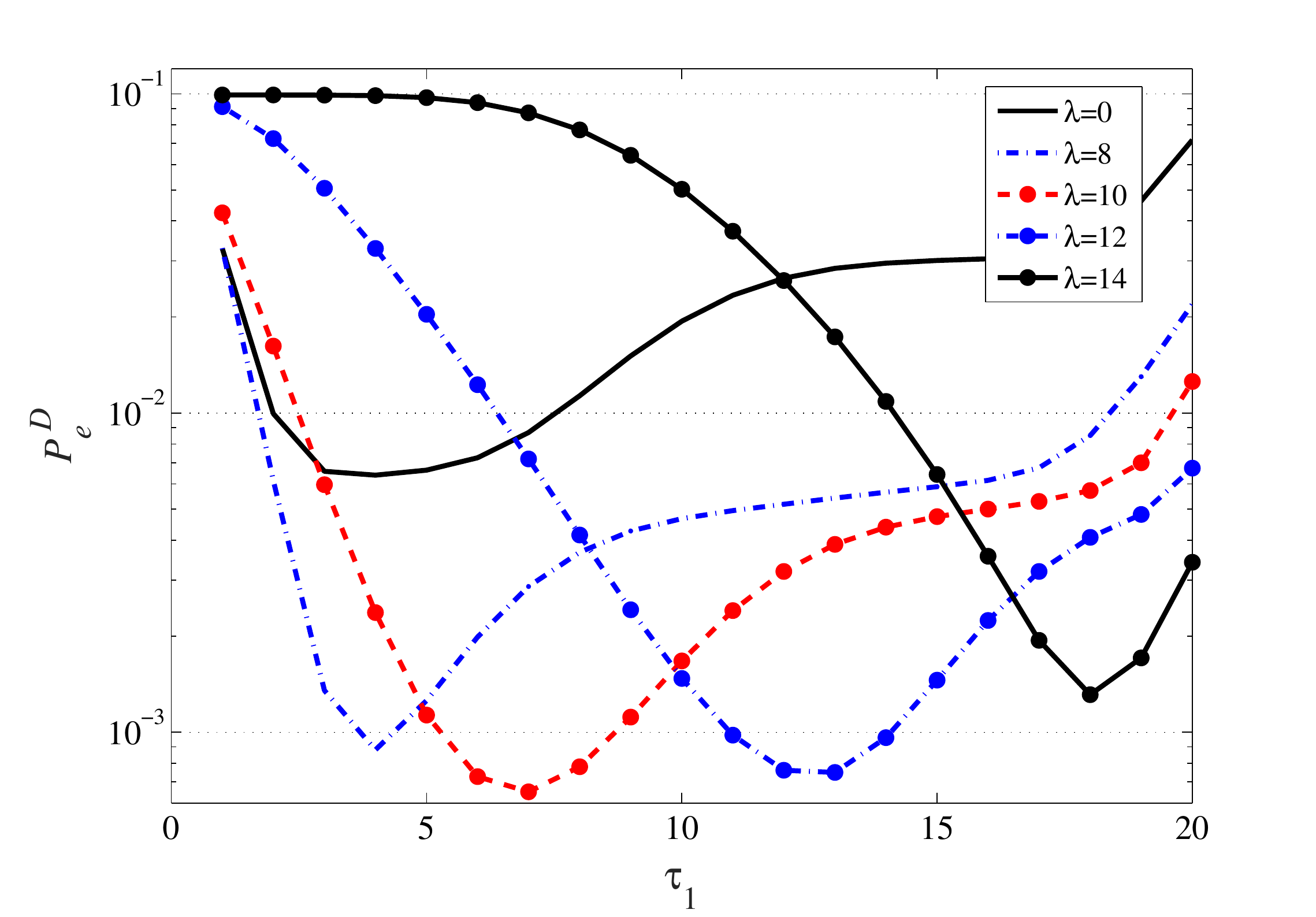}
		\vspace{-0.5em}
		\caption{The probability of detection error for the memoryless sensors versus threshold of FC.}
		\label{fig: Pe_thr}
		\vspace{0em}
	\end{figure}

	First, we present the results related to the abnormality detection.
	The detection error probability $P_e^\text{D}$ for memoryless sensors is shown versus the threshold $\tau_1$, in Fig. \ref{fig: Pe_thr}, for different marker noise densities $\lambda$.
	As can be seen, there is an optimal threshold $\tau_1^{\text{opt}}$, where the error probability is minimum, \emph{i.e.}, $P^{\text{D,opt}}_e$. It is also observed that the $\tau_1^{\text{opt}}$ increases for higher marker noise densities, due to the increase in the number of activated sensors and as a result the FC needs a larger $\tau_1$ for less detection error. The $P^{\text{D,opt}}_e$ for $\lambda =10$ is less than other densities. Because for smaller $\lambda$s the $P^{\text{D}}_{e|H_1}$ in \eqref{Eq: Pe} is dominant, which is decreasing function of $\lambda$, and for larger $\lambda$s the $P^{\text{D}}_{e|H_0}$ is dominant, which is increasing function~of~$\lambda$.
	
\begin{figure}
		\vspace{-1em}
		\centering
		\includegraphics[height=138pt]{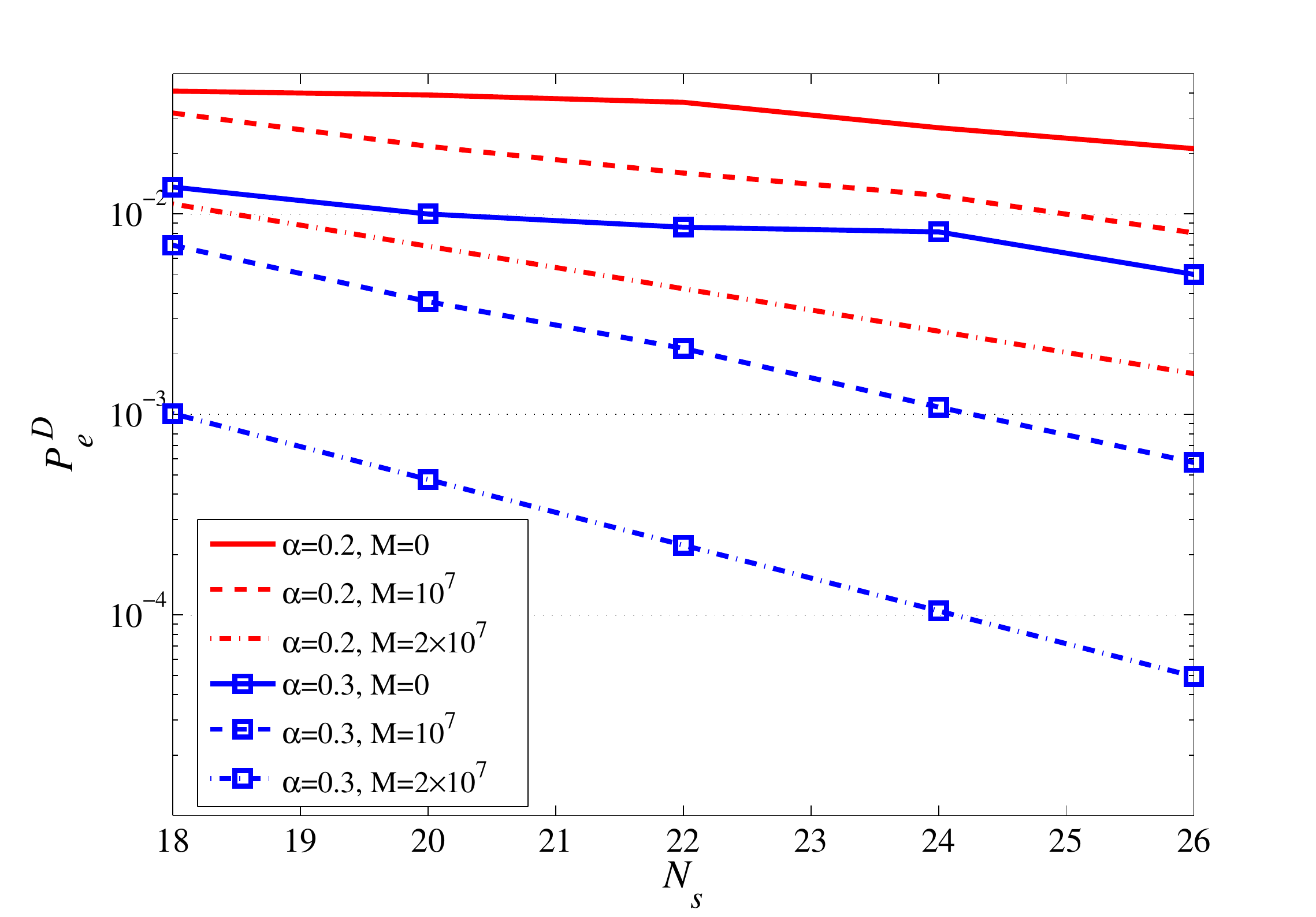}
		\vspace{-0.5em}
		\caption{The minimum probability of detection error versus the number of memoryless sensors.
		}
		\label{fig:M_alpha_1}
	\end{figure}

	In Fig. \ref{fig:M_alpha_1}, the minimum probability of detection error is plotted versus the number of sensors for $M=0, 10^7, 2 \times 10^7 \text{ and } \alpha=0.2, 0.3$. The $P^{\text{D}}_e$ has been  minimized over $\tau_1$ and $\tau_2$. The number of available molecules and activation probability $\alpha$ at the abnormality point are factors of strength and accuracy of the sensors.
	When $M=0$, the sensors do not activate each other and do not cooperate. It can be seen that the cooperation of sensors ($M > 0$) significantly improves the system performance in terms of the probability of error. Also it is observed that the minimum error probability decreases for higher number of markers and larger sensor activation probability $\alpha$ at the abnormality point. The weaker sensors (with smaller $\alpha$) need more cooperation with each other to decrease the probability of detection error.
	
\begin{figure}
		\vspace{-1em}
		\centering
		\includegraphics[height=138pt]{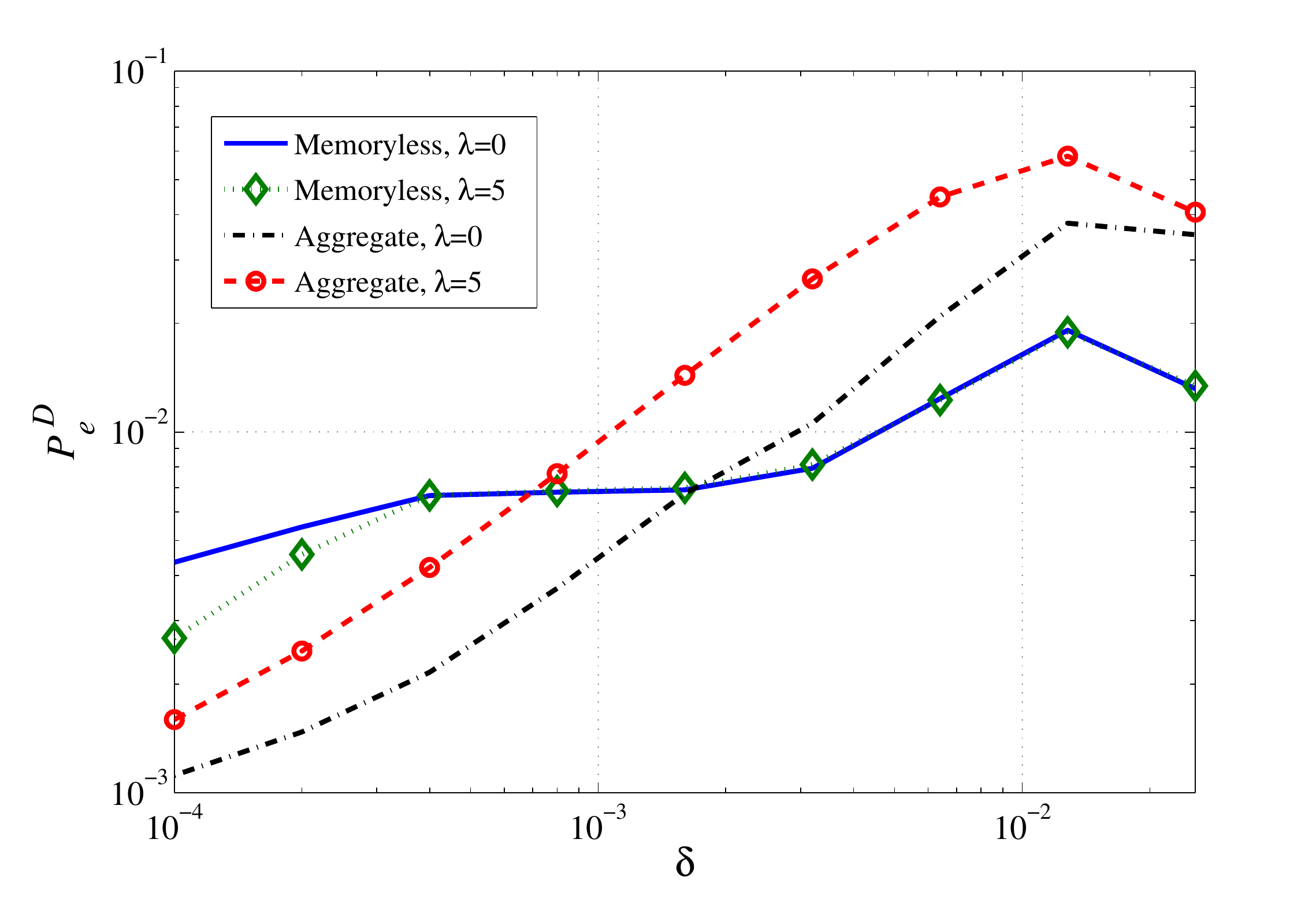}
		\vspace{-0.5em}
		\caption{The probability of detection error versus sensor noise for memoryless and aggregate sensors for $M=10^{8}$.}
		\label{fig:MemoAggre}
	\end{figure}

	In Fig. \ref{fig:MemoAggre}, we compare the performance of memoryless and aggregate sensors versus the sensor noise $\delta$. It can be seen that for small values of $\delta$, the aggregate sensors perform better than the memoryless ones. Because if sensor noise is weak, the $P^{\text{D}}_{e|H_1}$ is the dominant term (\emph{i.e.}, $P^{\text{D}}_{e|H_0} \ll P^{\text{D}}_{e|H_1}$) and aggregate sensors decrease the $P^{\text{D}}_{e|H_1}$ by aggregating the number of markers in all sampling times and amplifying the signal. For higher values of $\delta$, the $P^{\text{D}}_{e|H_0}$ is the dominant term and aggregate sensors increase the $P^{\text{D}}_{e|H_0}$ by amplifying the marker noise and the markers released by sensors that are activated by sensor noise.
	It can also be seen that by a small increase of the marker noise, the aggregate sensors, which consider the summation of observations, are influenced by stronger marker noise. Therefore, the error probability of aggregate sensors increases by increasing $\lambda$. For the memoryless sensors, which consider each of the observations separately until possible activation, the effect of marker noise is weak, without considerable affect on the error probability. Only when $\delta$ is very low, the $P^\text{D}_{e|H_1}$ is the dominant term and marker noise in fact helps the detection, and hereby $P^\text{D}_{e}$ decreases.
	
	Now, we present the localization results. To decrease the software running time, here we consider the number of sensors to be $N_s=10$ and
	show the effect of different parameters on the system performance for two considered cases of perfect and imperfect sensing (\emph{i.e.}, $\delta=0$ and $\delta>0$, respectively).
	
\begin{figure}
		\vspace{-1em}
		\centering
		\includegraphics[height=138pt]{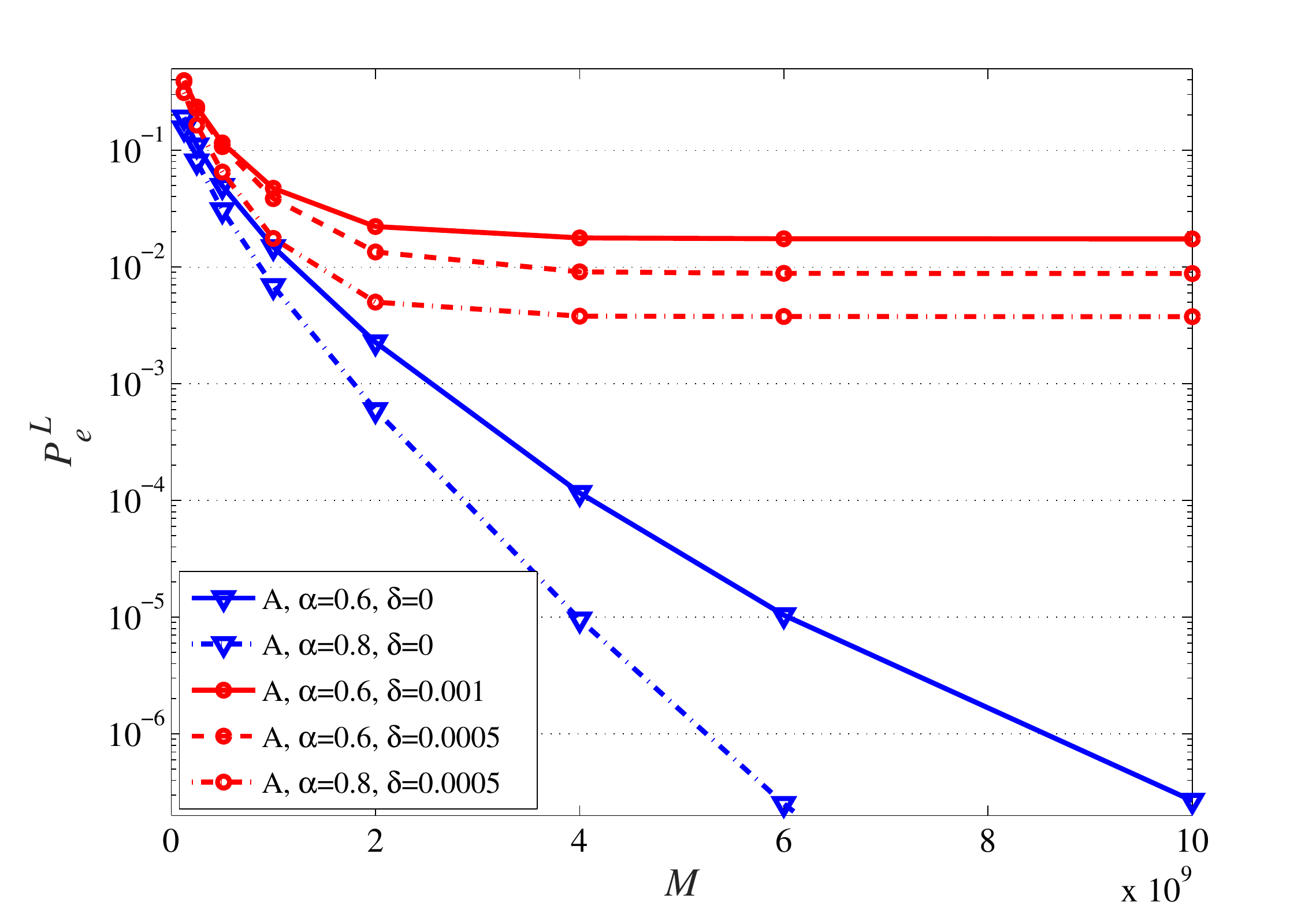}
		\vspace{-0.5em}
		\caption{The probability of localization error versus the number of released markers.}
		\label{fig:M_Loc}
	\end{figure}

	The probability of localization error versus the number of released markers $M$ is plotted in Fig. \ref{fig:M_Loc}, for type-A FCs in perfect and imperfect sensing regimes. As can be seen, for the perfect sensing case, if we increase the number of sensors, the probability of localization error decreases exponentially ($P_{e}^{\text{L},A} \rightarrow 0$).
	Also for the imperfect sensing case, the performance of type-A improves if we increase the number of markers. But if $M$ is large enough, its effect saturates and the probability of localization error does not change considerably ($P_{e}^{\text{L},A,\delta}>0$), which confirms Lemma \ref{Lemma_1}. Moreover, it can be seen that if $\delta$ increases, the probability of localization error increases too.
	Note that for perfect sensing case, the type-B detects the abnormality location with no error and for imperfect sensing case, it does not use the number of markers for its decision on abnormality location. Thus, $P_{e}^{\text{L},B,\delta}$ is constant versus $M$ and is not shown in Fig. \ref{fig:M_Loc}.
	
	\begin{figure}
		\vspace{-1em}
		\centering
		\includegraphics[height=138pt]{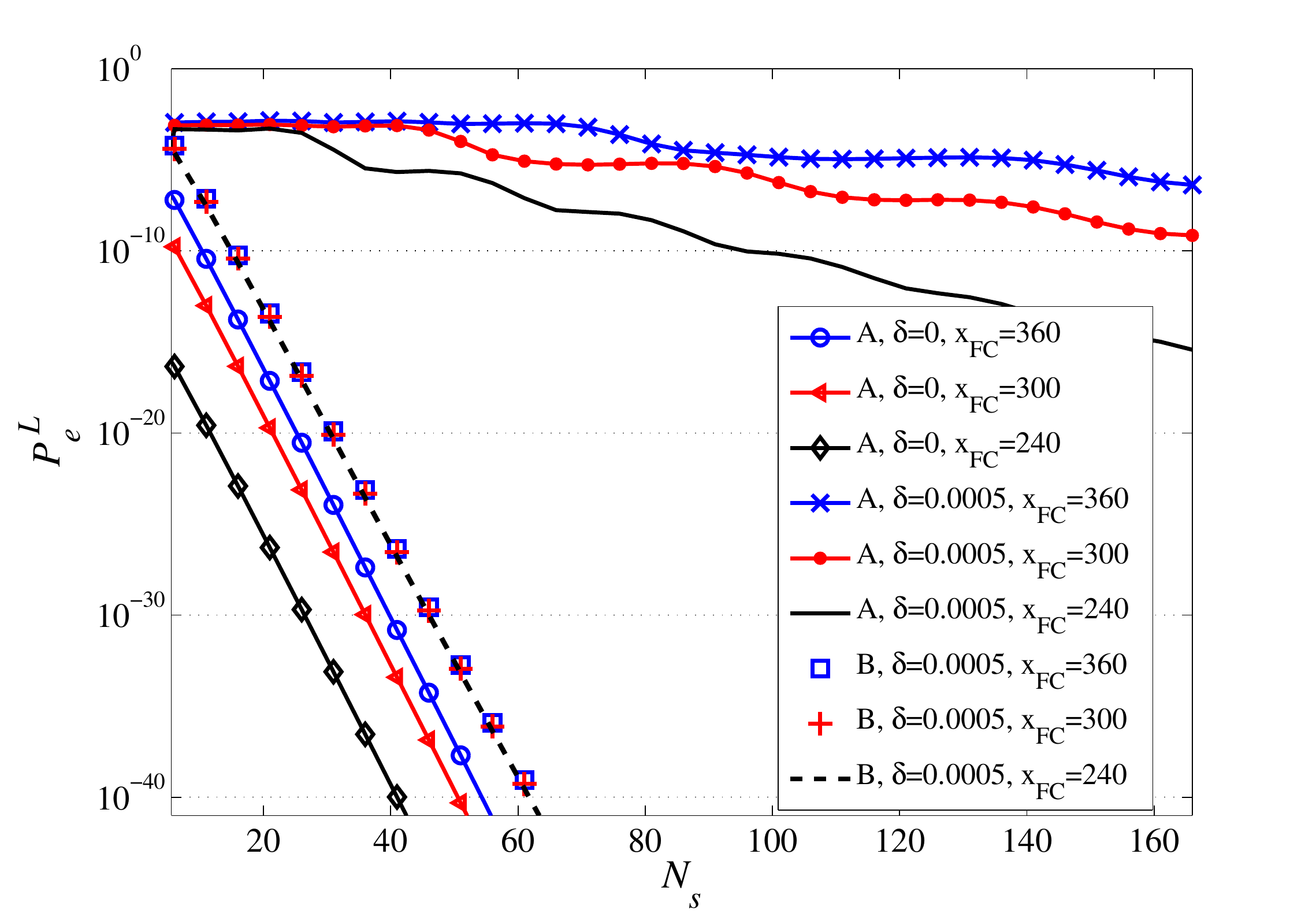}
		\vspace{-0.5em}
		\caption{The probability of localization error versus the number of sensors for $M=10^{10}$, $\alpha=0.8$.}
		\label{fig:Ns_Loc}
	\end{figure}
	\begin{figure}
		\centering
		\vspace{-1em}
		\includegraphics[height=138pt]{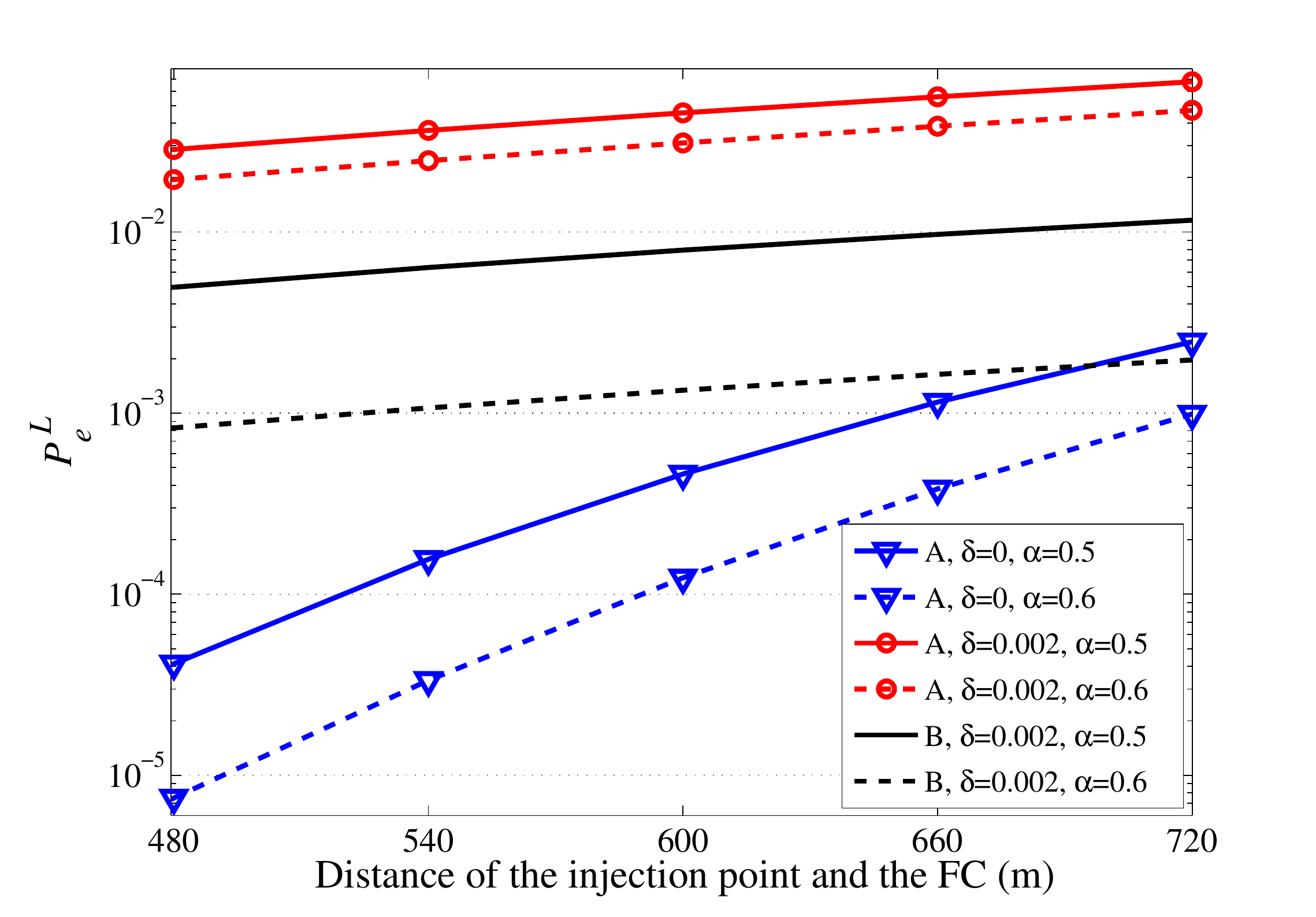}
		\vspace{-0.5em}
		\caption{The probability of localization error versus the distance of the injection point and the FC for $N_s=10,M=10^{10}$. Note that if $\delta=0$, we have $P_e^{\text{L}, B}=0$.}
		\label{fig:K_Loc}
	\end{figure}

	The probability of localization error versus the number of sensors $N_s$ is plotted in Fig. \ref{fig:Ns_Loc}, for perfect and imperfect sensing assumptions. As can be seen, if $\delta=0$ (perfect sensing), the probability of localization error for type-A FC decreases exponentially versus $N_s$ (note that the probability of localization error for type-B FC in perfect sensing regime is zero and thus is not shown). If $\delta >0$ (imperfect sensing), the effect of the number of sensors on the performance of type-A FC (\emph{{i.e.}, }$P_{e}^{\text{L},A,\delta}$) is less visible compared to the perfect sensing regime.
	Because in the imperfect sensing regime, some sensors may be activated at normal subregions and release markers there, which results in localization error. It is also seen that for larger $x_\text{FC}$, the $P_{e}^{\text{L},A,\delta}$ is increased. Because, the sensors have more time to be activated by the sensor noise, before they arrive the FC. In addition, the number of subregions is increased. Thus, the number of location hypotheses is increased too, which contributes further to higher error.
	The error probability of type-B FC decreases exponentially by increasing $N_s$ even for $\delta >0$.

	The probability of localization error versus the distance of the injection point and the FC (\emph{i.e.}, ($x_\text{FC}-x_0$)) is plotted in Fig.~\ref{fig:K_Loc}. As can be seen, the probabilities of localization error for both the perfect and imperfect sensing regimes are increasing versus the distance.
	But the effect of distance on the perfect sensing regime is more obvious than the imperfect sensing regime. According to Section \ref{Perfect sensing _type B}, the probability of localization error for type-B FC in the perfect sensing regime ($\delta =0 $) is zero.
	
	\section{Conclusion}
	\label{Sec: conclusion}
	In this paper, we have studied the concept of using MC-based cooperative system to detect an abnormality and to find its location in a cylindrical fluidic medium. For this purpose, we heve proposed a theoretical framework and analyzed the system performance metrics. Our model requires the mobile sensors to search the environment and then be absorbed to an FC. The sensors inform others by releasing markers in the environment after sensing the abnormality. Other sensors can be activated upon receiving these markers. The released markers were also used by the FC for the localization.
	For the purpose of localization, two types of FCs have been considered with different capabilities in specifying the sensors' storage levels.
	We have analyzed the performance of this system in terms of probability of error and showed that the cooperation among sensors in activating each other significantly improves the detection performance. Also we showed that for the localization problem, the performance is improved by using more sensors and markers and also by the increased FC resolution of the sensors' storage levels. In this paper, we assumed that at most one abnormality may occur in each region and considered the resolution of localization up to one subregion. Enhancing this resolution by exploiting the continuous sensors' storage levels and also studying multi-abnormality detection problem can be considered as future research directions. Furthermore, we have studied the laminar flow condition (due to its tractable analysis).  The effects of turbulent flow conditions and channel non-linearities on the system performance can be considered as another interesting and challenging future work. Also, future practical implementations are required to investigate the effect of further non-idealities.


	\begin{appendices}
		\section{Proof of Lemma \ref{Lemma_1}}\label{proof of lemma 1 and 2}
		The $P(R|J^*)$ in \eqref{Pe_new A,delta} is positive. Therefore, to have a vanishing $P_e^{\text{L}, A, \delta}$ when $M \rightarrow \infty$, we must have $P_{e|R,J^*=j}^{\text{L}, A, \delta} \rightarrow 0 , \forall R$ in \eqref{Pe_loc_B_delta}.
		If $M \rightarrow \infty$, from \eqref{gama_opt} we have $\gamma^{*}_{i}={M\sum_{i=1}^{K} r_i}\sqrt{ \mu'_{J^* K}\mu'_{(J^*+1) K}}$, and by assuming abnormality subregion $J^*=j$, the $P_{e|R,J^*=j}^{\text{L}, A, \delta}$ in \eqref{Pe_loc_B_delta} will be
		\begin{align}
		P_{e|R,J^*=j}^{\text{L}, A, \delta}=
		&
		1-Q(\frac{\sqrt{M}\sum_{i=1}^{K} r_i(\sqrt{  \mu'_{(j-1) K}\mu'_{j K}}-\mu'_{iK})}{\sqrt{\sum_{i=1}^{K} r_i {\mu'_{iK}}}})  \nonumber \\	&
		+Q(\frac{\sqrt{M} { \sum_{i=1}^{K} r_i( \sqrt{\mu'_{j K}\mu'_{(j+1) K}}}-\mu'_{iK})}{\sqrt{\sum_{i=1}^{K} r_i {\mu'_{iK}}}}). \nonumber
		\end{align}
		The $P_{e|R,J^*=j}^{\text{L}, A, \delta}$ in the above equation is zero only if the arguments of the first and the second Q-functions approach $-\infty$ and $+\infty$, respectively. This occurs when we have
		\begin{align}
		\sum_{i=1}^{K} r_i\sqrt{\mu'_{(j-1) K}\mu'_{j K}}  <  \sum_{i=1}^{K} r_i\mu'_{iK}  <  \sum_{i=1}^{K} r_i\sqrt{\mu'_{j K}\mu'_{(j+1) K}},
		\label{lem_1_cond_1}
		\end{align}
		which must be satisfied for any $R$. Substituting vector realizations of $r^{(s)}=[r_i^{(s)}]$, where $r_s^{(s)}\geq 1, r_i^{(s)}=0, \forall i\neq s$, in \eqref{lem_1_cond_1} results in
		\begin{align}
		\sqrt{\mu'_{(j-1) K}\mu'_{j K}}\overset{(\text{a})}{<}\mu'_{sK}\overset{(\text{b})}{<}\sqrt{\mu'_{j K}\mu'_{(j+1) K}}.
		\label{lem_1_cond_2}
		\end{align}
		From \eqref{MU equation}, we know $\mu'_{(i-1) K}<\mu'_{i K}, \forall i$ (note that $\mu'_{ij}$ is proportional to $\mu_{ij}$). Thus, the inequalities (a) and (b) in \eqref{lem_1_cond_2} are satisfied for $s\geq j$ and $s \leq j$, respectively. As a result, \eqref{lem_1_cond_2} is valid only for $s=j$ and vector realizations of $r^{(j)}=[0,\cdots,0,r^{(j)}_j\geq 1,0,\cdots,0 ]$. This means that all of the directly activated sensors are  activated at the abnormality subregion $J^*=j$, which is true only for the perfect sensing regime. Therefore, if $M \rightarrow \infty$ the probability of localization error for type-A FC tends to zero for perfect sensing regime, while it does not approach to zero for imperfect sensing regime.

		\section{Proof of Theorem \ref{Theorem_1}}\label{Appx: ML_type B}
		Assume that $\hat{J}^*=m$. Based on \eqref{ML_total_type_B} the following two conditions are true.
		\begin{align}
		\forall n>m&: C_1\overset{\Delta}{=}(1+\frac{\alpha}{\delta})^{r_{m}-r_n}(1-\frac{\alpha}{1-\delta})^{\sum_{i=m+1}^{n}r_i} > 1, \label{first_condition}\\
		\forall n<m&: C_2\overset{\Delta}{=}(1+\frac{\alpha}{\delta})^{r_{m}-r_n}(1-\frac{\alpha}{1-\delta})^{-\sum_{i=n+1}^{m}r_i} > 1.\label{second_condition}
		\end{align}
		We prove that $m = \max\underset{n=1,\cdots,K}{\operatorname{argmax}} \text{ } r_n$. We use contradiction. Thus, the following three cases may occur.
		\\
		$\bullet$ Case 1: $\exists n_1>m: r_{n_1}>r_m$. Substituting $n$ by $n_1$ in the left hand side (LHS) of \eqref{first_condition}, we have:
		\begin{align}
		&(1+\frac{\alpha}{\delta})^{r_{m}-r_{n_1}}(1-\frac{\alpha}{1-\delta})^{\sum_{i=m+1}^{n_1}r_i} \nonumber\\
		&\leq
		(1+\frac{\alpha}{\delta})^{-1}(1-\frac{\alpha}{1-\delta})^{\sum_{i=m+1}^{n_1}r_i} \nonumber\\
		&\overset{\text{(a)}}{<} (1-\frac{\alpha}{1-\delta})^{N_s-1+\sum_{i=m+1}^{n_1}r_i}  <  1, \label{not_case1}
		\end{align}
		where (a) follows from condition $(1+\frac{\alpha}{\delta})(1-\frac{\alpha}{1-\delta})^{N_s-1}>1$.
		We see that \eqref{not_case1} is in contradiction with \eqref{first_condition}.
		\\
		$\bullet$ Case 2: $\exists n_1<m: r_{n_1}>r_m$. Substitute $n$ by $n_1$ in the LHS of \eqref{second_condition}, we obtain:
		\begin{align}
		&(1+\frac{\alpha}{\delta})^{r_{m}-r_{n_1}}(1-\frac{\alpha}{1-\delta})^{-\sum_{i=m+1}^{n_1}r_i} \nonumber\\
		&\leq
		(1+\frac{\alpha}{\delta})^{-1}(1-\frac{\alpha}{1-\delta})^{-\sum_{i=m+1}^{n_1}r_i} \nonumber\\
		&\overset{\text{(a)}}{<} (1-\frac{\alpha}{1-\delta})^{N_s-1-\sum_{i=m+1}^{n_1}r_i}  \overset{\text{(b)}}{<}  1, \label{not_case2}
		\end{align}
		where (a) follows from condition $(1+\frac{\alpha}{\delta})(1-\frac{\alpha}{1-\delta})^{N_s-1}>1$ and (b) follows from $N_s-1\geq \sum_{i=m+1}^{n_1}r_i$.
		Again, we observe that \eqref{not_case2} is in contradiction with \eqref{second_condition}.
		\\
		$\bullet$ Case 3: $\exists n_1>m: r_{n_1}=r_m$. Substituting $n$ by $n_1$ in the LHS of \eqref{first_condition}, we have:
		\begin{align}
		&(1-\frac{\alpha}{1-\delta})^{\sum_{i=m+1}^{n_1}r_i}   <  1, \label{not_case3}
		\end{align}
		and we see that \eqref{not_case3} is in contradiction with \eqref{second_condition}. Thus, if $(1+\frac{\alpha}{\delta})(1-\frac{\alpha}{1-\delta})^{N_s-1}>1$, we have $m = \max\underset{n=1,\cdots,K}{\operatorname{argmax}} \text{ } r_n$.
		
	\end{appendices}
	
	\section*{Acknowledgment}
	
	The authors would like to thank Dr. Amin Gohari for his helpful suggestions and comments.
	

	\begin{IEEEbiographynophoto}{Ladan Khaloopour}
		received the B.Sc. and M.Sc. degrees in electrical engineering from Sharif University of Technology, Tehran, Iran, in
		2015 and 2017, respectively. She is currently pursuing
		the Ph.D. degree in electrical engineering at
		Sharif University of Technology. Her research interests
		include molecular and wireless communications, information theory and communication networks.
	\end{IEEEbiographynophoto}
	
	\begin{IEEEbiographynophoto}{Mahtab Mirmohseni}
		received the B.Sc., M.Sc., and Ph.D. degrees in communication systems from the Department of Electrical Engineering, Sharif University of Technology, Iran, in 2005, 2007, and 2012, respectively, where she has been an Assistant Professor with the Department of Electrical Engineering since Spring 2014. She is also affiliated with the Information Systems and Security Laboratory, Sharif University of Technology. She was a Postdoctoral Researcher with the School of Electrical Engineering, Royal Institute of Technology, Sweden, till February 2014. Her current research interests include different aspects of information theory, mostly focusing on molecular communication and secure and private communication. She was a recipient of the Award of the National Festival of the Women and Science (Maryam Mirzakhani Award) in 2019 and also was selected as an Exemplary Reviewer of the IEEE Transactions on Communications in 2016. Her current research interests include different aspects of information theory, mostly focusing on molecular communication and secure and private communication.
	\end{IEEEbiographynophoto}
	
	\begin{IEEEbiographynophoto}{Masoumeh Nasiri-Kenari}
		received the B.Sc. and M.Sc. degrees in electrical engineering from the Isfahan University of Technology, Isfahan, Iran, in 1986 and 1987, respectively, and the Ph.D. degree in electrical engineering from The University of Utah, Salt Lake City, UT, USA, in 1993. From 1987 to 1988, she was a Technical Instructor and a Research Assistant with the Isfahan University of Technology. Since 1994, she has been with the Department of Electrical Engineering, Sharif University of Technology, Tehran, Iran, where she is currently a Professor. Dr. Nasiri-Kenari founded the Wireless Research Laboratory (WRL), Electrical Engineering Department, in 2001, to coordinate the research activities in the field of wireless communication. Current main activities of WRL are energy harvesting and green communications, 5G, and molecular communications. From 1999 to 2001, she was a Co-Director of the Advanced Communication Research Laboratory, Iran Telecommunication Research Center, Tehran. She was a recipient of the 2014 Premium Award for the Best Paper in IET Communications. She received the Distinguished Researcher Award and the Distinguished Lecturer Award from the EE Department, Sharif University of Technology, in 2005 and 2007, respectively, the Research Chair on Nano Communication Networks from the Iran National Science Foundation, and the Research Grant on Green Communication in Multi-Relay Wireless Networks from Swedish Research Council during 2015–-2018. Since 2014, she has been serving as an Associate Editor for the IEEE Transactions on Communications.
	\end{IEEEbiographynophoto}
	
\end{document}